\documentclass[12pt]{article}
\usepackage{latexsym}
\usepackage[dvipdfmx]{graphicx}
\usepackage{amsmath}
\usepackage[bookmarks,bookmarksnumbered]{hyperref}
\usepackage[top=3.5cm,bottom=3.5cm,left=2cm,right=2cm]{geometry}
\usepackage{pifont}          
\usepackage{amsmath,amssymb}
\usepackage{color}
\usepackage{geometry}
\usepackage{caption}
\usepackage{cancel}
\newcommand{\nn}{\nonumber\\}
\newcommand{\h}{\hspace}
\newcommand{\be}{\begin{equation}}
\newcommand{\e}{\end{equation}}
\newcommand{\aln}[1]{\begin{align}#1\end{align}}
\newcommand{\paren}[1]{\left(#1\right)}
\newcommand{\p}{\partial}
\newcommand{\sqbr}[1]{\left[#1\right]}
\newcommand{\br}[1]{\left\{#1\right\}}
\begin{document}
\title{
\vspace{-2cm}
\vbox{
\baselineskip 14pt
\hfill \hbox{\normalsize KEK-TH-2233}} 
\vskip 1cm
\bf \Large   On Preheating in Higgs Inflation 
\vskip 0.5cm
}
\author{
Yuta Hamada$^{a}$\thanks{E-mail: \tt hamada(at)apc.in2p3.fr},\h{1mm} 
 Kiyoharu~Kawana$^{bc}$\thanks{E-mail: \tt kawana(at)post.kek.jp}\h{1mm} and 
 Adam Scherlis$^{d}$\thanks{
 E-mail: \tt adam(at)scherlis.com
 }
\bigskip\\
\it 
\normalsize
 $^a$ Universit\'e de Paris, CNRS, Astroparticule et Cosmologie, F-75006 Paris, France,\\
 \normalsize
\it  $^b$ Theory Center, High Energy Accelerator Research Organization (KEK),\\
\normalsize
\it
 $^c$ Center for Theoretical Physics, Department of Physics and Astronomy,\\
\it
\normalsize 
 Seoul National University, Seoul 08826, Korea, \\ 
 \normalsize
\it  $^d$ Stanford Institute for Theoretical Physics, Department of Physics,
\\
 \normalsize
\it Stanford University, Stanford, CA 94305, United States \\
\smallskip
}
\date{\today}

\maketitle

\abstract{
Recently, the problem of unitarity violation during the preheating stage of Higgs inflation with a large non-minimal coupling has been much discussed in the literature. 
We point out that this problem can be translated into a strong coupling problem for the dimensionless effective coupling, and that the existence of these problems is highly dependent on the choice of higher-dimensional operators because they can significantly change the background dynamics and the canonical normalization of the fluctuations around it.       
Correspondingly, the typical energy of particles produced during the first stage of preheating can remain comparable to or below the cutoff scale of the theory.  
As an example, we numerically calculate the particle production in the presence of a specific four-derivative operator of the Higgs field, and confirm the statement above. 
Our argument also applies to multi-field inflation with non-minimal couplings.
}
\newpage

\section{Introduction}
Higgs inflation \cite{Bezrukov:2007ep,Rubio:2018ogq,Steinwachs:2009zz} with non-minimal gravitational coupling between the Ricci scalar and the Higgs is one of the simplest and most natural scenarios for cosmological inflation because it can be realized within the Standard Model (SM).\footnote{
It is known that the SM Higgs potential can have a saddle point around the Planck scale when the top mass is around $171$ GeV. 
However, we cannot use this fact to realize inflation purely within the SM because its inflationary predictions are inconsistent with cosmological observations \cite{Isidori:2007vm,Hamada:2013mya,Fairbairn:2014nxa}. 
Therefore, we need some extensions or mechanisms in order to regard the Higgs as an inflaton. 
Non-minimal coupling is one such possible extension. 
} 
Higgs inflation is also connected to interesting phenomenology such as the production of primordial black holes~\cite{Ezquiaga:2017fvi,Bezrukov:2017dyv,Pi:2017gih,Rasanen:2018fom,Cheong:2019vzl}.
In the case of a quartic potential i.e. $V(H)= \lambda_H (H^\dagger H)^2$, the quartic coupling $\lambda_H$ and the non-minimal coupling $\xi$ are constrained by the cosmic microwave background (CMB) normalization as $\lambda_H/\xi^2\sim 10^{-9}$. 
Thus, as long as $\lambda_H={\cal{O}}(10^{-3} \text{--} 10^{-2})$, $\xi$ has to have an unnaturally large value ${\cal{O}}(10^{3\text{--}4})$. Such a large value of $\xi$ causes a few physical inconsistencies.     
For example, the unitarity of the model has been viewed with suspicion \cite{Burgess:2009ea,Barbon:2009ya,Barvinsky:2009ii,Burgess:2010zq,Bezrukov:2010jz,Giudice:2010ka} because the naive tree-level cutoff scale of the model is given by $\Lambda^{}:=M_{pl}^{}/\xi$, and this is comparable with or smaller than other typical energy scales during the inflation. 
If this is the case, we cannot neglect higher-dimensional contributions such as 
$(H^\dagger H)^n/\Lambda^{2n-4}\ (n\geq 3)$ 
which is dominant over the quartic potential in the large-field region. 
Thus, the predictability or consistency of the model seems to be (strongly) UV-dependent. 
However, this problem was solved in \cite{Bezrukov:2010jz} where it is argued that the cutoff scale $\Lambda$ is a background-dependent quantity and can become sufficiently large relative to the relevant dynamical scales during the inflation.

However, large $\xi$ (or small cutoff scale) is still problematic\footnote{It is argued that this problem is absent in Palatini Higgs inflation \cite{Rubio:2019ypq,Karam:2020rpa}.} when it comes to the preheating stage after inflation \cite{Ema:2016dny,DeCross:2015uza,DeCross:2016fdz,DeCross:2016cbs,Sfakianakis:2018lzf}, see also \cite{Bezrukov:2008ut,GarciaBellido:2008ab} for the earlier studies of (p)reheating after Higgs inflation.
As discussed in \cite{Ema:2016dny}, the background dynamics of the inflaton shows some spike-like behavior around its zero-crossings and can cause violent particle production of the Nambu-Goldstone (NG) modes of the inflaton or the longitudinal modes of the weak gauge boson. 
In particular, the typical energy scale of these produced particles is ${\cal{O}}(\sqrt{\lambda_H}M_{pl}^{})\gg \Lambda$, so the consistency of the theory during particle production is not clear.   
Therefore, even if it is consistent during inflation, the preheating dynamics still indicates the necessity of extending this model to a UV model which preserves unitarity.  

In this paper, as a first step toward understanding the dynamics of these unitary models, we analyze the problem of unitarity during particle production after Higgs inflation by taking the effects of higher-dimensional operators into account. 
Among various higher-dimensional operators involving $H$, the most important ones are those that include $\partial H$ because these problems originate in the behavior of $\partial H$ around its (first) zero-crossing.   
We show that the issue of the strong coupling does not happen if operators like $\paren{|\p_\mu H|^2}^n/\Lambda^{4(n-1)}$ are present because they change the definition of the canonical field in terms of $H$ around the zero-crossings. 
We study the dynamics of $H$ and particle production after inflation in the presence of a specific choice of higher-dimensional operator i.e. $\paren{|\p_\mu H|^2}^2/\Lambda^{4}$. 
We confirm that the strong coupling is absent in this case.

On the other hand, if an operator such as $(|H|^2)^2\paren{|\p_\mu H|^2}^2/\Lambda^{8}$ is added in addition to $\paren{|\p_\mu H|^2}^2/\Lambda^{4}$, we again face the strong coupling problem in general. 
Therefore, the fate of the perturbativity is highly UV-dependent, and the result changes depending on what types of higher dimensional operators are considered. 
From the low energy field theory point of view, there are infinitely many possibilities for the higher dimensional operators, and it is impossible to fix them.\footnote{The asymptotic scale/shift symmetry~\cite{Bezrukov:2010jz} does not constrain the operators relevant here.}
See \cite{Hamada:2016onh} for the related discussion of the difference between prescription I and II in Higgs inflation.
We claim that, among the infinitely many possibilities, there exist choices where the strong coupling does not arise, and the analysis of the preheating of Higgs inflation is self-consistent.
\\

This paper is organized as follows. 
In Section \ref{sec:review}, we briefly review inflation with a non-minimal coupling. 
In Section \ref{sec:preheating}, we first show that the unitarity violation problem is translated into the strong coupling problem of the dimensionless effective couplings. 
Then, we show that the unitarity or strong coupling problem during the preheating stage is highly dependent on the choice of higher-dimensional operators in subsection \ref{sec:higher}.  
Next, we study particle production after inflation in the presence of a specific but important operator i.e. $|\partial_\mu^{} H|^4/\Lambda^4$ in subsection \ref{Sec:particle_prodution2}. 
As is expected from the resolution of the unitarity problem, the existence of such a term significantly reduces the typical energy scale of produced particles, and we will see that this reduction is marginally consistent with the cutoff scale of the theory. 
The summary is presented in Section \ref{sec:Summary}.
In \ref{app:unitarity}, we briefly review the unitarity issue of Higgs inflation.
The background dynamics in the conventional Higgs inflation is reviewed in \ref{app:dynamics in conventional case}. 
The detail calculation of the particle production of the NG mode is presented in \ref{app:NG}. 
\\

\section{Inflation with Non-minimal Coupling}\label{sec:review}
In order to fix our notation, we first briefly review Higgs inflation \cite{Bezrukov:2007ep}. 
See also \cite{Rubio:2018ogq} for the review. 
In the following discussion, the Higgs field in the Jordan frame is denoted by $H$, and the Hubble parameter in the Einstein frame is represented by ${\cal{H}}$ to distinguish it from the Higgs field.   

We start from the following action in the Jordan frame:
\aln{
S=&\int d^4x\sqrt{-g_J^{}}\left( \frac{1}{2}M_{pl}^2\Omega^2 R_J^{}
- g_J^{\mu\nu} D_\mu^{} H \left(D_\nu^{} H\right)^* - V_J^{}(H) + \cdots \right),
}
where 
\aln{
V_J(H)=\lambda_H \left(|H|^2\right)^2,
\quad \Omega^2=1 + 2\xi \frac{|H|^2}{M_{pl}^2}.  
\label{eq: Einstein frame potential}
}
By performing the following redefinition of the metric field,
\aln{ 
g_{\mu\nu}^{}=\Omega^2 g_{J\mu\nu}^{},
}
we get
\aln{
R_J=\Omega^2
\sqbr{
R_E
+3\Box\log \Omega^2 
-{3\over 2}g^{\mu\nu}{(\p_\mu \log \Omega^2)(\p_\nu \log \Omega^2)}
},
}
which leads to the action in the Einstein frame:
\aln{
S=&\int d^4x\sqrt{-g}\left(\frac{1}{2}M_{pl}^2R
- {1\over \Omega^2} g^{\mu\nu} D_\mu^{} H \left(D_\nu^{} H\right)^*
-\frac{3}{4} M_{pl}^2 (\p_\mu \log \Omega^2)^2
-V(H) + \cdots \right),
\label{eq:action in Einstein frame}
}
where 
\aln{
V(H) := \frac{V_J^{}(H)}{\Omega^4}
}
is the potential in the Einstein frame. For the field value $|H|^2 \gg M_{pl}^2/\xi$, the kinetic term for the radial component of Higgs field is effectively given by the third term (the second term is neglected). Therefore, the canonically normalized field is
\aln{
\chi:= \sqrt{3\over2} M_{pl} \log \Omega^2\quad  (\text{for}\ |H|^2 \gg M_{pl}^2/\xi).
}
Then, the action is
\aln{
S=&\int d^4x\sqrt{-g}\sqbr{\frac{1}{2}M_{pl}^2R
- {1\over \Omega^2} g^{\mu\nu} D_\mu^{} H \left(D_\nu^{} H\right)^*
-\frac{1}{2} (\p_\mu \chi)^2
- {\lambda_H\over4} {M_{pl}^4\over \xi^2} \paren{1-e^{-\sqrt{2\over3}{\chi\over M_{pl}}}}^2+ \cdots },
}
from which we can see that the height of the potential is ${\cal O}(\lambda_H^{}M_{pl}^4/\xi^2)$. 
Correspondingly, the Hubble scale is
\aln{{\cal H}=\frac{1}{M_{pl}}\sqrt{\frac{V}{3}}\sim \frac{1}{2}\sqrt{\frac{\lambda_H^{}}{3}}\frac{M_{pl}^{}}{\xi}. 
}
The slow roll parameters are calculated as 
\aln{
\epsilon=\frac{M_{pl}^2}{2}\left(\frac{V'}{V}\right)^2\simeq \frac{4}{3}
\exp\left(-2\sqrt{\frac{2}{3}}\frac{\chi}{M_{pl}^{}}\right),\quad \eta=M_{pl}^2\frac{V^{''}}{V^{}}
\simeq -{4\over3}\exp\left(-\sqrt{\frac{2}{3}}\frac{\chi}{M_{pl}^{}}\right),
} 
where the prime represents the derivative with respect to $\chi$.
In addition, we also have a relation between the e-folding number $N$ and $\epsilon,\eta$:
 \aln{
N=\int dt H\simeq\frac{1}{M_{pl}^2}\int d\varphi \frac{V}{V^{'}}\simeq \frac{\sqrt{3}}{2\epsilon^{1/2}}\quad  
\Rightarrow
\quad\epsilon=\frac{3}{4N^2},\quad \eta=-\frac{1}{N}. 
\label{eq:relation between N and epsilon}
}
Note that $\lambda_H^{}$ and $\xi$ are not completely independent parameters because they are constrained by the CMB normalization~\cite{Akrami:2018odb} 
\aln{
A_s^{}:= \frac{V^{}}{24\pi^2\epsilon M _{pl}^4}\bigg|_{k=k_* ^{}}^{}\simeq2.2\times 10^{-9}.
\label{eq:CMB}
}  
where $k_*^{}\simeq0.05\text{Mpc}^{-1}$ is the pivot scale. 
By substituting Eq.~(\ref{eq:relation between N and epsilon}) into Eq.~(\ref{eq:CMB}), we obtain 
\aln{
\frac{\lambda_H}{\xi^2}\simeq6.3\times\left(\frac{50}{N}\right)^2\times 10^{-10}. 
\label{eq:lambdaH_xi}
} 
The value of $\lambda_H$ is typically ${\cal{O}}(10^{-2})$ for a smaller top quark mass which realizes a stable electroweak vacuum~\cite{Degrassi:2012ry,Hamada:2012bp,Buttazzo:2013uya}. In this case, the corresponding value of the non-minimal coupling is $\xi={\cal{O}}(10^{3\text{--}4})$. If $\lambda_H$ is tuned to be small at inflationary energy scales,\footnote{There are several proposals for an underlying mechanism of this tuning~\cite{Froggatt:1995rt,Shaposhnikov:2009pv,Hamada:2015dja,Eichhorn:2017als}.} then $\xi$ can be as small as ${\cal{O}}(10)$, which is known as the critical Higgs inflation scenario~\cite{Hamada:2014iga,Bezrukov:2014bra,Hamada:2014wna}.

Throughout this paper, we consider the case $\xi={\cal{O}}(10^{3\text{--}4})$ and leave the analysis of (p)reheating in critical Higgs inflation for future investigation.

\section{Unitarity Issue at the Preheating Stage in Higgs Inflation
}
\label{sec:preheating}
\subsection{Strong coupling problem during preheating}
Let us consider the dynamics after inflation. 
As mentioned in the Introduction and \ref{app:unitarity}, the unitarity of this model during inflation is maintained because the cutoff scale is field-dependent \cite{Burgess:2009ea,Barbon:2009ya,Burgess:2010zq,Bezrukov:2010jz,Giudice:2010ka}. 
On the other hand, it is known that there is still a unitarity issue during the preheating stage~\cite{Ema:2016dny,DeCross:2015uza,DeCross:2016fdz,DeCross:2016cbs,Sfakianakis:2018lzf}. 
Here, we discuss this problem from a different point of view than these references, in the context of the strong coupling of the Higgs field. 
This viewpoint makes it easier to understand the role of higher-dimensional operators in unitarizing the theory in the next subsection.  

Our starting point is Eq.~\eqref{eq:action in Einstein frame}.  
If we ignore the gauge field, the action is
\aln{
S=&\int d^4x\sqrt{-g}\left(\frac{1}{2}M_{pl}^2R
- {1\over \Omega^2} |\p_\mu^{} H|^2 
-\frac{3}{\Omega^4} {\xi^2 \over M_{pl}^2} (\p_\mu |H|^2)^2
-V(H) + \cdots \right).
\label{eq:action in Einstein frame 2}
}
We would like to consider the zero-crossing of the Higgs field. As we will see, the typical momentum scale associated with this process is much larger than the Hubble scale, and we can safely regard the background geometry as flat. Moreover, we can take $\Omega^2$ to be $1$ because the Higgs field is close to the origin. This results in the simplified action
\aln{
S=&\int d^4x\sqrt{-g}\sqbr{\frac{1}{2}M_{pl}^2R
- |\p_\mu^{} H|^2 
- {3 \xi^2 \over M_{pl}^2} (\p_\mu |H|^2)^2
-V(H) + \cdots }.
\label{eq:simplified action}
}
Let us parametrize $H$ as
\aln{\label{eq:parametrization}
H={1\over\sqrt{2}}
\begin{pmatrix}
\phi_1 + i \phi_2 \\
\phi_3 + i \phi_4 \\
\end{pmatrix}.
}
By using this parametrization, the third term in Eq.~\eqref{eq:simplified action} contains the $\phi_i^2 (\p \phi_j)^2$ terms.
For $i\neq j$, this is understood as the interaction term while, for $i=j$, 
this term changes the canonical normalization of $\phi_i^{}$ field. 
To discuss whether the system is suffering from a strong coupling problem, it is convenient to move to the canonically normalized frame of the Higgs fields.   

Indeed, we can move to the canonical frame order-by-order in $|H|^2$, by redefining the Higgs fields as\footnote{It is possible to do the redefinition exactly, but it is not necessary here.}
\aln{
H \to H - {2\xi^2\over M_{pl}^2} |H|^2 H + {6\xi^4\over M_{pl}^4} (|H|^2)^2 H+\cdots,
}
which leads to
\aln{
S=&\int d^4x\sqrt{-g} \bigg[\frac{1}{2}M_{pl}^2R
- |\p_\mu^{} H|^2 
+ {\xi^2 \over M_{pl}^2} 
\bigg\{ 4 |\p_\mu H|^2 |H|^2 - (\p_\mu |H|^2)^2 \bigg\}
\nn&\phantom{int d^4x\sqrt{-g} \bigg[}
+{2\xi^4\over M_{pl}^4} \br{ -4 |\p_\mu H|^2 (|H|^2)^2 + |H|^2  \paren{\p_\mu |H|^2}^2 }
-\tilde{V}(H)
+ \cdots \bigg],
\label{eq:simplified action 2}
}
where
\aln{
\tilde{V}(H):= V\paren{H - {2\xi^2\over M_{pl}^2} |H|^2 H + {6\xi^4\over M_{pl}^4} (|H|^2)^2 H}
=\lambda_H (|H|^2)^2
-8 \lambda_H {\xi^2 \over M_{pl}^2}(|H|^2)^3
+\cdots.
}
We can see that the expressions in the curly brackets in Eq.~\eqref{eq:simplified action 2} do not contain the term $\phi_i^4$ and $\phi_i^6$ with two derivatives, as expected.
On the other hand, other terms cannot be removed by the field redefinition, and can cause strong couplings as we will show in the following. 

Next, suppose that the inflaton component is $\phi_1^{}$. 
Then, the third term in Eq.~\eqref{eq:simplified action 2} provides effective mass terms for $\phi_{2,3,4}$,
\aln{\label{eq:effective_mass}
\mathcal{L}_{\phi^2}&={\xi^2 \over M_{pl}^2} 
\bigg\{ 4 |\p_\mu H|^2 |H|^2 - (\p_\mu |H|^2)^2 \bigg\}
\nn&=-{\lambda_H M_{pl}^2 \over 10} \paren{\dot{\phi}_1^2\over 0.1 \lambda_H M_{pl}^4/\xi^2} \paren{\phi_2^2 + \phi_3^2 + \phi_4^2}+\cdots,
}
where we focused on the region $\phi_1^{}\sim0, \dot{\phi}_1^2 \sim 0.1\lambda_H M_{pl}^4/\xi^2$ corresponding to the first zero-crossing. 
See \ref{app:dynamics in conventional case} for details. 
The fourth term in Eq.~\eqref{eq:simplified action 2} induces the following effective four-field terms,
\aln{
\mathcal{L}_{\phi^4}&={2\xi^4\over M_{pl}^4} \br{ -4 |\p_\mu H|^2 (|H|^2)^2 + |H|^2  \paren{\p_\mu |H|^2}^2 }
\nn&=
{2\xi^4\over M_{pl}^4} \dot{\phi}_1^2 \bigg\{ 2(|H|^2)^2 -  |H|^2 \phi_1^2 \bigg\}+\cdots
\nn&= {\lambda_H \xi^2 \over 10} \paren{ \dot{\phi}_1^2 \over 0.1 \lambda_H M_{pl}^4/\xi^2} 
\paren{\phi_1^2+\phi_2^2+\phi_3^2+\phi_4^2} \paren{\phi_2^2+\phi_3^2+\phi_4^2}+\cdots
\nn&=:-{1\over4} \kappa \phi_1^2 \paren{\phi_2^2 + \phi_3^2 + \phi_4^2} 
- {1\over4!}\rho \paren{\phi_2^4 + \phi_3^4 + \phi_4^4} 
+\cdots,
}
where
\aln{
&\kappa= -{2\lambda_H \xi^2 \over 5} \paren{ \dot{\phi}_1^2 \over 0.1 \lambda_H M_{pl}^4/\xi^2},
&&\rho=-{12\lambda_H \xi^2 \over 5} \paren{ \dot{\phi}_1^2 \over 0.1 \lambda_H M_{pl}^4/\xi^2}.
}
One can see that if $\lambda_H \xi^2 \gg1$, the effective $\phi^4$ couplings are non-perturbative, and the controllability of the theory is lost.\footnote{For $\lambda_H \xi^2 \gg1$, the perturbativity is broken when $\dot{\phi}_1^2\gtrsim (M_{pl}/\xi)^4$. This is interpreted as the unitarity violation scale, see Eq.~\eqref{eq:delH4}. } 
This coupling is perturbative as long as the condition 
\aln{
&\lambda_H \,\xi^2 \lesssim 10\pi \paren{ \dot{\phi}_1^2 \over 0.1 \lambda_H M_{pl}^4/\xi^2}^{-1} \paren{\kappa_{\text{max}}\over 4\pi},
\quad
{5\over3}\pi \paren{ \dot{\phi}_1^2 \over 0.1 \lambda_H M_{pl}^4/\xi^2}^{-1} \paren{\rho_{\text{max}}\over 4\pi}
}
is satisfied. 
Here, $\kappa_{\text{max}}$ and $\rho_{\text{max}}$ are the maximum allowable values of the couplings $\kappa$ and $\rho$, whose exact values are subject to debate. 
Combined with Eq.~\eqref{eq:lambdaH_xi}, we have 
\aln{\label{eq:xi_bound}
\xi \lesssim 470\, \sqrt{N \over 50} \paren{ \dot{\phi}_1^2 \over 0.1 \lambda_H M_{pl}^4/\xi^2}^{-{1\over4}} \paren{\kappa_{\text{max}}\over 4\pi}^{1\over4},
\quad
300\, \sqrt{N \over 50} \paren{ \dot{\phi}_1^2 \over 0.1 \lambda_H M_{pl}^4/\xi^2}^{-{1\over4}} \paren{\rho_{\text{max}}\over 4\pi}^{1\over4}.
}
In terms of $\lambda_H$, the bound is
\aln{\label{eq:lambda_bound}
\lambda_H \lesssim
1.4\times10^{-4} \paren{50\over N}\paren{ \dot{\phi}_1^2 \over 0.1 \lambda_H M_{pl}^4/\xi^2}^{-{1\over2}} \paren{\kappa_{\text{max}}\over 4\pi}^{1\over2}
,\quad
5.7\times 10^{-5} \paren{50\over N}\paren{ \dot{\phi}_1^2 \over 0.1 \lambda_H M_{pl}^4/\xi^2}^{-{1\over2}} \paren{\rho_{\text{max}}\over 4\pi}^{1\over2}. 
}

The bounds Eqs.~\eqref{eq:xi_bound}\eqref{eq:lambda_bound} are also applicable to multi-field inflation with non-minimal couplings.
We can see that the perturbativity condition is not satisfied in the conventional Higgs inflation scenario \cite{Bezrukov:2007ep} while the critical Higgs inflation scenario \cite{Hamada:2014iga,Bezrukov:2014bra,Hamada:2014wna} does not suffer from this problem. 
Nevertheless, if one naively continues the calculation of particle production assuming that the tree level action is correct, one finds very efficient production of the NG modes or the longitudinal gauge boson~\cite{Ema:2016dny,DeCross:2015uza,DeCross:2016fdz,DeCross:2016cbs,Sfakianakis:2018lzf}. 

The values obtained in Eq.~\eqref{eq:xi_bound} are roughly consistent with the value $\xi^2\lesssim 5\times10^4$ in \cite{Sfakianakis:2018lzf}, where the bound was derived by requiring that the maximum excited wavenumber is less than the unitarity violation scale $M_{pl}^{}/\xi$.

\subsection{Effect of Higher Dimensional Operators}
\label{sec:higher}
Per the discussion in the previous subsection, it is impossible to study the preheating dynamics from a low-energy point of view. 
However, in this subsection, we show that for some specific choices of the higher-dimensional operators, the strong coupling issue does not arise.
As an example, let us consider 
\aln{\label{eq:delH4}
& {c \over \Lambda_J^4} \paren{|\p_\mu H|^2}^2,
&& \Lambda_J:= \frac{M_{pl}\Omega}{\xi}\left(1 + c' \left(\frac{\xi}{M_{pl}\Omega}\right)^2|H|^2\right)
\sim {M_{pl}\over \xi} \quad \text{for $|H|^2\ll {M_{pl}^2\over\xi^2}$}
}
where $c$ and $c'$ are the coefficients, and $\Lambda_J$ is the cutoff scale in Jordan frame estimated in Ref.~\cite{Bezrukov:2010jz}.\footnote{Compared with \cite{Bezrukov:2010jz}, we treat the coefficient $c'$ as a free parameter.} 
Note that this operator multiplied by $\sqrt{-g}$ is invariant under the transformation $g_{\mu\nu}\to \Omega^{-2} g_{\mu\nu}$ and does not change in the Einstein frame. 
When the Higgs crosses zero, this operator gives the effective kinetic term for the Higgs as 
\aln{
-c{\xi^4 \over2 M_{pl}^4}\dot{\phi}_1^2 |\p_\mu H|^2, 
}
where $\dot{\phi}_1^2$ is almost constant (of order given in Eq.~\eqref{eq:derivative with O2}). 
Hence, for the canonically normalized field $\tilde{H}:= \sqrt{c/2}\,\xi^2\,|\dot{\phi}_1| H/M_{pl}^2$, the effective mass Eq.~\eqref{eq:effective_mass} becomes smaller, and the four point coupling Eq.~\eqref{eq:simplified action 2} remains perturbative.
We notice that Eq.~\eqref{eq:delH4} generates an additional contribution to the effective mass and coupling. 
By expanding $\Lambda_J^{-4}$ as a function of $|H|^2$, we obtain
\aln{\label{Eq:new_term}
\Delta\mathcal{L}_{\phi^2}&
=-\sqrt{\frac{\lambda_H}{c}}\frac{c'}{\xi} M_{pl}^2 \frac{\dot{\phi}_1^2}{\sqrt{\frac{\lambda_H}{c}}\frac{M_{pl}^4}{\xi^3}}\sum_i\tilde{\phi}_i^2,
\nonumber\\
\Delta\mathcal{L}_{\phi^4}&=\frac{5c'^2}{2c}\left(\sum_i\tilde{\phi}_i^2\right)^2,
}
which are again perturbative for moderate values of $c$ and $c'$. 
In this way, the problem of the strong coupling is avoided, and we can safely discuss the dynamics of the preheating.

More generally, we may consider the operators 
\aln{\label{eq:general_operators}
&c_n \sqbr{{1 \over \Lambda_J^4} |\p_\mu H|^2}^n |\p_\mu H|^2, 
\quad d_m \sqbr{{1 \over \Lambda_J^4} |\p_\mu H|^2}^m (|H|^2)^2, 
}
where the first operator reduces the effective coupling by changing the kinetic term while the second one leads to a large coupling. In order to avoid the problem of strong coupling, the condition $2n\geq m$ needs to be realized.\footnote{We assume that the coefficient of the $\dot{\phi}_1^{2n+2}$ term in the Hamiltonian is positive. For $n=1$, this follows from causality, unitarity, and analyticity arguments in a broad class of the theory \cite{Adams:2006sv}.}  
A similar perturbative condition is required for the other terms in the SM lagrangian such as the Yukawa and gauge couplings, and the Higgs mass term as well. 

From the low energy viewpoint, there are infinitely many choices of $c_n, d_m$.
Computing these values is a purely UV question. 
Here we point out that, among infinitely many possibilities from the low energy viewpoint, there exist choices where the self-consistency of the preheating process of Higgs inflation is maintained.

One may think that an operator such as Eq.~\eqref{eq:delH4} is inevitably generated by loop corrections, and that the problem of unitarity violation would disappear thanks to the loop-induced higher derivative term.   
However, since the loop correction accompanies many other operators, we also leave this possibility for future investigation.

During the inflation, the operator \eqref{eq:delH4} may not modify the classical dynamics of the Higgs field but may play the role to generate the quantum fluctuation of the Higgs field. It is interesting to study if there is any observable consequence such as non-Gaussianities.

\section{Particle Production
}\label{Sec:particle_prodution2}
In this subsection, we discuss particle production with a specific choice of the higher-dimensional operator i.e. Eq.~\eqref{eq:delH4}. 
We first study the background dynamics of the inflaton $\phi_1^{}$ in the presence of Eq.~\eqref{eq:delH4}. 
See \ref{app:dynamics in conventional case} (and Ref.~\cite{Ema:2016dny}) for a discussion of the dynamics without higher-dimensional operators. 
The existence of Eq.~\eqref{eq:delH4} weakens the spike-like behavior in the inflaton dynamics \cite{Ema:2016dny}. 
As a result, the amount of particle production and the typical energy scale of the produced particles are reduced.  

\subsection{Background Dynamics with Higher Dimensional operator}\label{subsection:dynamics with HDO}
In the presence of Eq.~\eqref{eq:delH4}, the equation of motion of $\phi_1^{}$ is
\aln{\label{eq:eom_w_cutoff}
&\ddot{\phi}_1^{}\sqbr{1+ {3c\over\Lambda_J^4} \dot{\phi}_1^2 \paren{d\phi_1\over d\chi}^2}+3{\cal H}\dot{\phi}_1^{}\sqbr{1+ {c\over \Lambda_J^4} \,\dot{\phi}_1^2 \paren{d\phi_1\over d\chi}^2}+\frac{d^2\chi}{d\phi_1^2}\frac{d\phi_1^{}}{d\chi}\dot{\phi}_1^2
-3\left(\frac{d\phi_1^{}}{d\chi}\right)^2\frac{c\frac{{\partial\Lambda}_J}{\partial\phi_1}}{\Lambda_J^5}\dot{\phi}_1^4
+{\left(\frac{d\phi_1^{}}{d\chi}\right)^2}\frac{\partial V^{}}{\partial \phi_1^{}}=0,
}
where $\chi$ is the canonically normalized field of $\phi_1^{}$ whose exact definition is given by Eq.~\eqref{eq:relation between varphiJ and varphi} in \ref{app:dynamics in conventional case}.  
Around the origin, by neglecting the Hubble friction term and putting $d\chi/d\phi_1^{}\sim1,\ d^2\chi/d\phi_1^{2}\sim3\xi^2\phi_1^{}/M_{pl}^2$, this equation approximately becomes 
\aln{\label{eq:eom_w_cutoff_2}
\ddot{\phi}_1^{}+  \ddot{\phi}_1^{}{3 c \over \Lambda_J^4} \dot{\phi}_1^2+3\xi^2\frac{\phi_1^{}}{M_{pl}^2}\dot{\phi}_1^2+\lambda_H^{} \phi_1^{3}\sim 0,
} 
from which we can see that the typical time scale is $ {\cal{O}}(M_{pl}^{}/\xi)$ and 
\footnote{
The second and third terms in Eq.~\eqref{eq:eom_w_cutoff_2} cancel each other when $\ddot{\phi}\sim -(M_{pl}^{}/\xi)\phi_1^{}$. 
Higher derivative terms are determined by taking the derivative. 
}
\aln{
&{d^{2n}\phi_1 \over d t^{2n}}= {\cal{O}}\paren{\paren{M_{pl}\over\xi}^{2n} \phi_1^{}}, 
&&{d^{2n+1}\phi_1^{} \over d t^{2n+1}}= {\cal{O}}\paren{\paren{M_{pl}\over\xi}^{2n}\dot{\phi}_1^{}}, \quad \quad \text{for  $0\leq n\in \mathbb{Z}, \,\, \phi_1\ll {M_{pl} \over \xi}$}.
}
Note that the effects of Hubble friction do not change this parametric estimate because ${\cal{H}}={\cal{O}}(\sqrt{\lambda_H^{}}M_{pl}^{}/\xi)$.    
The value of $\dot{\phi}_1^{}$ around the origin can be evaluated by energy conservation, 
\aln{
{1\over2}\dot{\chi}^2+V+{c\over4\Lambda_J^4}\dot{\phi}_1^4\sim\frac{\lambda_H^{} M_{pl}^4}{4\xi^2},
}
from which we obtain
\aln{
|\dot{\phi}_1^{}|\sim {\lambda_H^{1/4}  \over \xi^{3/2} c^{1/4}} M_{pl}^2 
\sim \frac{1}{\paren{c\,\lambda_H \xi^2}^{1/4}}M_{pl}^{}{\cal{H}}.
\label{eq:derivative with O2} 
}
From this equation, we see that the spike-like feature is weaker than in the conventional case i.e. Eq.~(\ref{eq:phijdot2}) by a factor of $\paren{c\,\lambda_H \xi^2}^{-1/4}$. 
If we consider the higher order $c_n^{}$ operators in   Eq.~\eqref{eq:general_operators} instead of the four derivative operator, the value of $|\dot{\phi}_1^{}|$ becomes smaller. 
Specifically, the first operator in Eq.~\eqref{eq:general_operators} leads to a $\dot{\phi}_1^{2n+2}/\Lambda_J^{4n}$ term in the Lagrangian. 
By equating this to the inflation energy, we have 
\aln{\dot{\phi}_1^{}\sim (\lambda_H^{}\xi^2)^{\frac{1}{2n+2}}\Lambda_J^2
\sim \paren{\xi\over200}^{\frac{2}{n+1}}\paren{50\over N}^{\frac{1}{n+1}} \Lambda_J^2,
} 
where the CMB normalization condition Eq.~(\ref{eq:lambdaH_xi}) is used.  
This is a decreasing function of $n$ for $\xi\gtrsim 200$. 

Another important quantity for particle productions is the period for which the maximum value Eq.~\eqref{eq:derivative with O2} is maintained. We denote the period by $\Delta t_{\text{sp}}^{}$.
The peak momentum of produced particles is given by the mass scale $m_{\text{sp}}=1/\Delta t_{\text{sp}}^{}$. 
The time scale $\Delta t_{\text{sp}}^{}$ is evaluated by using energy conservation.  
The energy density is dominated by the $3\xi^2\phi_1^2(\partial\phi_1^{})^2/M_{pl}^2$ term just after inflation.
Then, the four-derivative term (\ref{eq:delH4}) becomes comparable to the $3\xi^2\phi_1^2(\partial\phi_1^{})^2/M_{pl}^2$ term at $\phi_1^{}=\tilde{\phi}$ where 
\aln{
\tilde{\phi}^2\sim \frac{\lambda_H^{} M_{pl}^2}{\xi}.  
\label{phitilde}
}
Here, we have used $\dot{\phi}_1^{2}\sim \lambda_H^{}M_{pl}^{4}/\xi^{3}$ right after inflation.  
(see Eq.~(\ref{eq:phijdot}) in \ref{app:dynamics in conventional case}). 
The time scale $\Delta t_{\text{sp}}^{}$ is estimated as
\aln{\frac{1}{m_{\text{sp}}}=\Delta t_{\text{sp}}^{}=\frac{
\tilde{\phi}}{\dot{\phi}}\sim (c\lambda_H^{})^{1/4}\frac{\xi}{M_{pl}^{}},
\label{eq:new time scale}
}
which is longer than the typical time scale in Higgs inflation without the four derivative operator, $\Delta t_{\text{sp}}^0\sim (\sqrt{\lambda_H^{}}M_{pl}^{})^{-1}$. (see Eq. (\ref{eq:spike scale 1}) in \ref{app:dynamics in conventional case}). 
The spike-like behavior of the inflaton becomes milder compared with the conventional case.  
It is expected that, if we add the higher order $c_n$ operator in Eq.~\eqref{eq:general_operators}, the spike-like behavior becomes much milder.
\begin{figure}[t]
\begin{center}
\begin{tabular}{c}
\includegraphics[width=8cm]{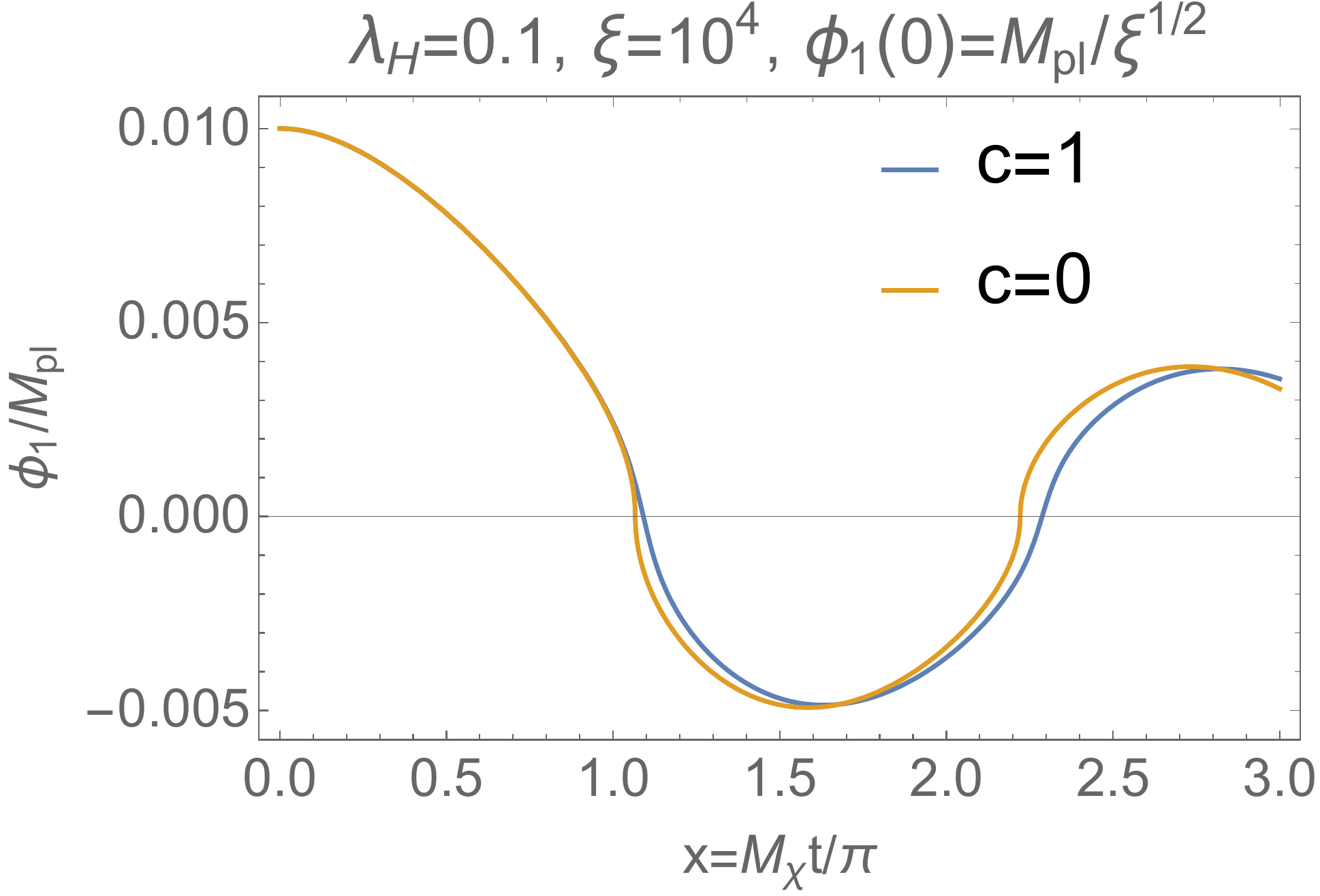}
\includegraphics[width=8cm]{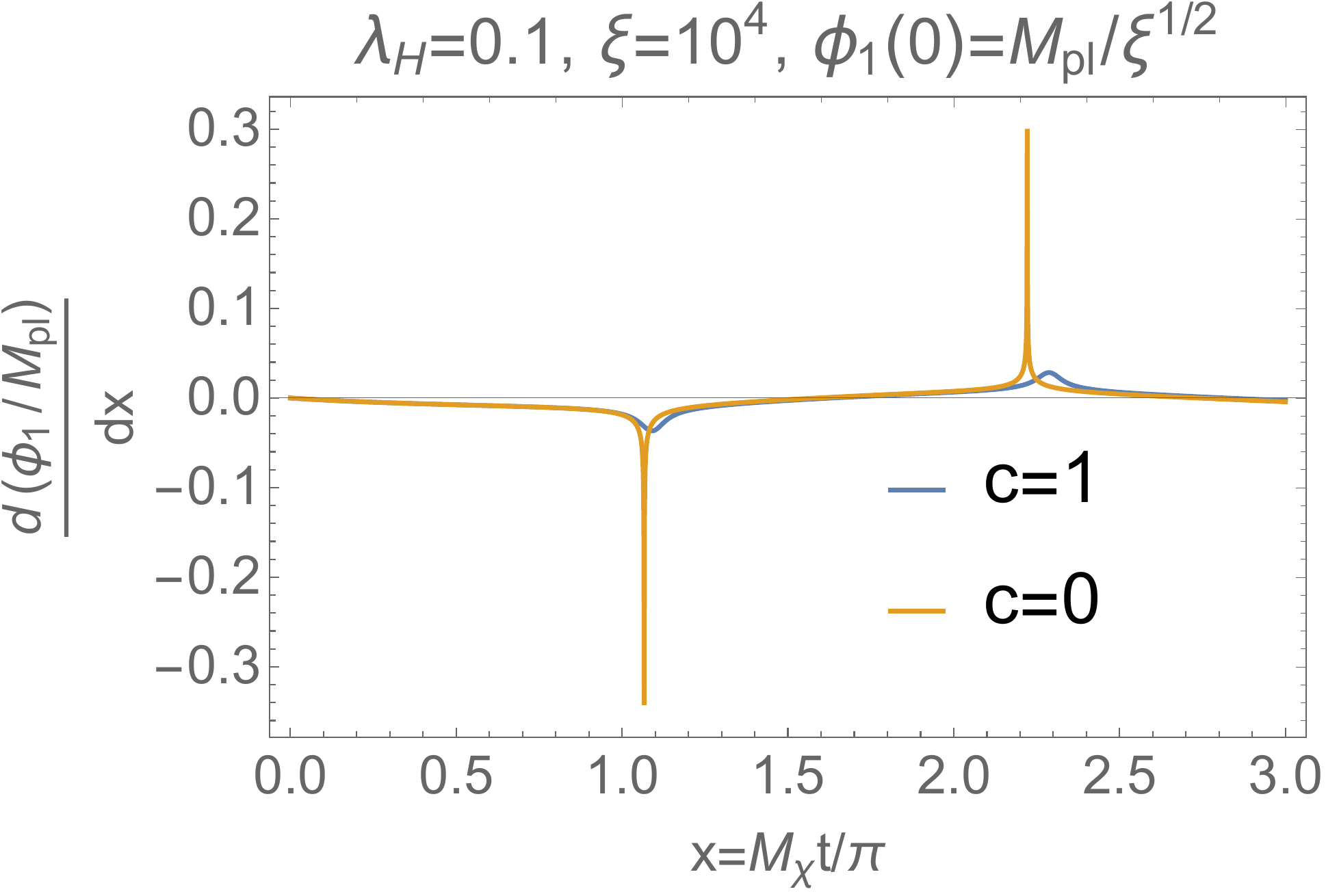}
\end{tabular}
\end{center}
\caption{Upper left: The time evolution of $\phi_1^{}$ for $c=0$ (orange) and $c=1$ (blue).
Upper right: The time evolution of $\dot{\phi}_1^{}$ for $c=0$ (orange) and $c=1$ (blue).
The parameter $c'$ is taken to be $0$. The unit of the time is Eq.~\eqref{Eq:M_chi}.
}
\label{fig:time evolution including O2}
\end{figure}

In Fig.~\ref{fig:time evolution including O2}, we show our numerical calculations of $\phi_1^{}$ (left) and $\dot{\phi}_1^{}$ (right), where blue and orange lines correspond to $c^{}=1$ and $0$, respectively.   
Here, we define
\aln{\label{Eq:M_chi}
M_\chi^{2}=\lambda_H^{}M_{pl}^2/(3\xi^2)
\simeq \paren{1.45\times 10^{-5}M_{pl}}^2 \paren{\frac{\lambda_H/\xi^2}{6.3\times10^{-10}}},
}
as a typical Higgs mass scale after inflation, 
and we choose $\lambda_H^{}=0.1,\ \xi=10^4$.  
We see that the maximum value of $|\dot{\phi}_1|$ with $c=1$ is suppressed by a factor of ${\cal O}(0.1)$ compared with that of $|\dot{\phi}_1|$ with $c=0$.  
Furthermore, the maximum value of $|\dot{\phi}_1|$ with $c=1$ is consistent with Eq.~(\ref{eq:derivative with O2}) within an order of magnitude. 

The spike-like behavior of $\dot{\phi}_1^{}(t)$ was first pointed out in \cite{Ema:2016dny}, and this property plays an important role in particle production after Higgs inflation because the mass of particles typically depends on the derivatives of $\phi_1^{}(t)$ when it is defined in the Jordan frame. 

\subsection{Brief Review of Particle Production}
Let us consider particle production. 
Because the dynamics of $\phi_1^{}$ is significantly changed, the amount of produced particles 
also changes. 
In particular, as a consequence of Eq.~(\ref{eq:new time scale}), the typical energy scale of the produced particles is reduced by a factor of ${\cal{O}}(\xi^{-1})$ compared with the conventional case, i.e. $m_{\text{sp}}^0$.  

The time evolution of particle number density of a particle species $\varphi$ is described by the Heisenberg equation~\cite{Kofman:1994rk,Kofman:1997yn,Ema:2016dny}:
\aln{\label{Eq:fluctuations}
&\varphi''_k + \omega_k^2 \varphi_k =0,  
&&\omega_k^2=k^2+m_\varphi^2,
}
with the initial conditions
\aln{
&\left.\varphi_k\right|_{\eta=0}=\left.\frac{1}{\sqrt{2\omega_k}}\right|_{\eta=0},
&&\left.\varphi'_k\right|_{\eta=0}=\left.-i\omega_k \varphi_k\right|_{\eta=0},
 }
Here, a prime is the derivative with respect to the conformal time $\eta$, $\varphi_k$ is the Fourier mode of the fluctuation $\varphi$, $k$ is the absolute value of the comoving momentum of $\varphi$, and $m_\varphi$ is the time-dependent mass. 
Equivalently, we can solve
\aln{\label{Eq:Bogiliubov}
&\alpha'_k(\eta)=\frac{\omega_k'}{2\omega_k}e^{2i\int^{\eta}_0d\eta'\omega_k(\eta')}\beta_k(\eta),
&&\beta'_k(\eta)=\frac{\omega_k'}{2\omega_k}e^{-2i\int^{\eta}_0d\eta'\omega_k(\eta')}\alpha_k(\eta),
&&\alpha_k(0)=1,
&\beta_k(0)=0,
}
where $\alpha$ and $\beta$ are the Bogiliubov coefficients:
\aln{
\varphi_k=\frac{1}{\sqrt{2\omega_k}}\paren{
\alpha_k e^{-i\int^\eta_0 d\eta'\omega_k(\eta')}+\beta_k e^{i\int^{\eta}_0\eta' \omega_k(\eta')}
}.
}
The particle number density $n_\varphi$ per physical volume  is defined by
\aln{ 
&n_\varphi(\eta)=\frac{1}{a^3} \int \frac{d^3k}{(2\pi)^3}f_\varphi(\eta,k), 
&& f_\varphi(\eta,k)=\frac{1}{2\omega_k}\paren{|\varphi_k'^2|+\omega_k^2|\varphi_k|^2}-\frac{1}{2}= |\beta_k|^2,
}
from which we define the energy density $\rho_\varphi$ per physical volume as
\aln{\label{Eq:energy_density}
&\rho_\varphi(\eta)=\frac{1}{a^4}\int_0^\infty d\paren{\log k}\tilde{\rho}_\varphi,
&&\tilde{\rho}_\varphi=\frac{k^3\omega_k}{2\pi^2}f_\varphi.
}
In the following, we numerically study particle production in the following two cases:
\begin{itemize}
\item The inflaton is the Higgs field of a global $U(1)$ symmetry. We consider the production of the massless NG boson. 
\item The inflaton is the Higgs field of a $U(1)$ gauge symmetry. We consider the production of the longitudinal mode of the gauge boson.
\end{itemize}
The second one is close to the Higgs inflation case, and is of primary interest. 
On the other hand, the first one is not directly related to Higgs inflation. 
However, since a statement which is analogous to the equivalence theorem holds, we will find that the production of the NG boson and longitudinal gauge boson are similar for the high-momentum region.

As benchmark points, we study
\aln{
&\lambda_H=10^{-3}\,\text{and}\,10^{-2},
&&c=1,
&&c'=0,\,0.1,\,\text{and} \,1.
}
The non-minimal coupling $\xi$ is chosen in such a way that Eq.~\eqref{eq:lambdaH_xi} is satisfied:
\aln{
\xi\simeq\begin{cases}
1260
\quad\text{for $\lambda_H=10^{-3}$}\\
3980
\quad\text{for $\lambda_H=10^{-2}$}
\end{cases}.
}
In the model with the $U(1)$ gauge symmetry, we take the gauge coupling $g$ to be $0.1$.

%
%
\subsection{Nambu-Goldstone Mode}
Let us consider the production of the NG mode $\theta(x)$ of a complex scalar $\varphi_J^{}(x)=r_J^{}(x)e^{i\theta_J^{}(x)/M_{pl}^{}}/\sqrt{2}$ where $r_J^{}(x)$ is regarded as an inflaton. 

After the Weyl transformation $g_{\mu\nu}^{}=\Omega^2 g_{J\mu\nu}^{}$, the action is 
\aln{
S&=\int d^4x\sqrt{-g}\left(-\frac{1}{2}g^{\mu\nu}\partial_\mu^{}r\partial_\nu^{}r-\frac{1}{2\Omega^2}g^{\mu\nu}\left(\frac{r_J^{}}{M_{pl}^{}}\right)^2\partial_\mu^{}\theta_J^{}\partial_\nu^{}\theta_J^{}-V(r)
+\frac{c}{\Lambda_J^4}(|\partial^\mu\varphi_J^{}|^2)^2
\right),
\label{eq:action of complex inflaton}
}
where $r$ is the canonically normalized field of $r_J^{}$. 
From this action, we obtain the effective mass of  the NG mode by defining the canonically normalized field $\theta(x)$. 
After the computation described in \ref{app:NG}, the frequency $\omega_k$ in Eq.~\eqref{Eq:fluctuations} is given by
\aln{
&\omega_k^2=k^2+m_\theta^2,
&&m_\theta^2=-{F''\over F},
&&
F={a r_J\over\Omega M_{pl}} \sqrt{1+ {c \,\Omega^2 \dot{r}_J^2 \over \Lambda_J^4}}.
\label{NG effective mass}
}

\begin{figure}[t]
\begin{center}
\includegraphics[width=8cm]{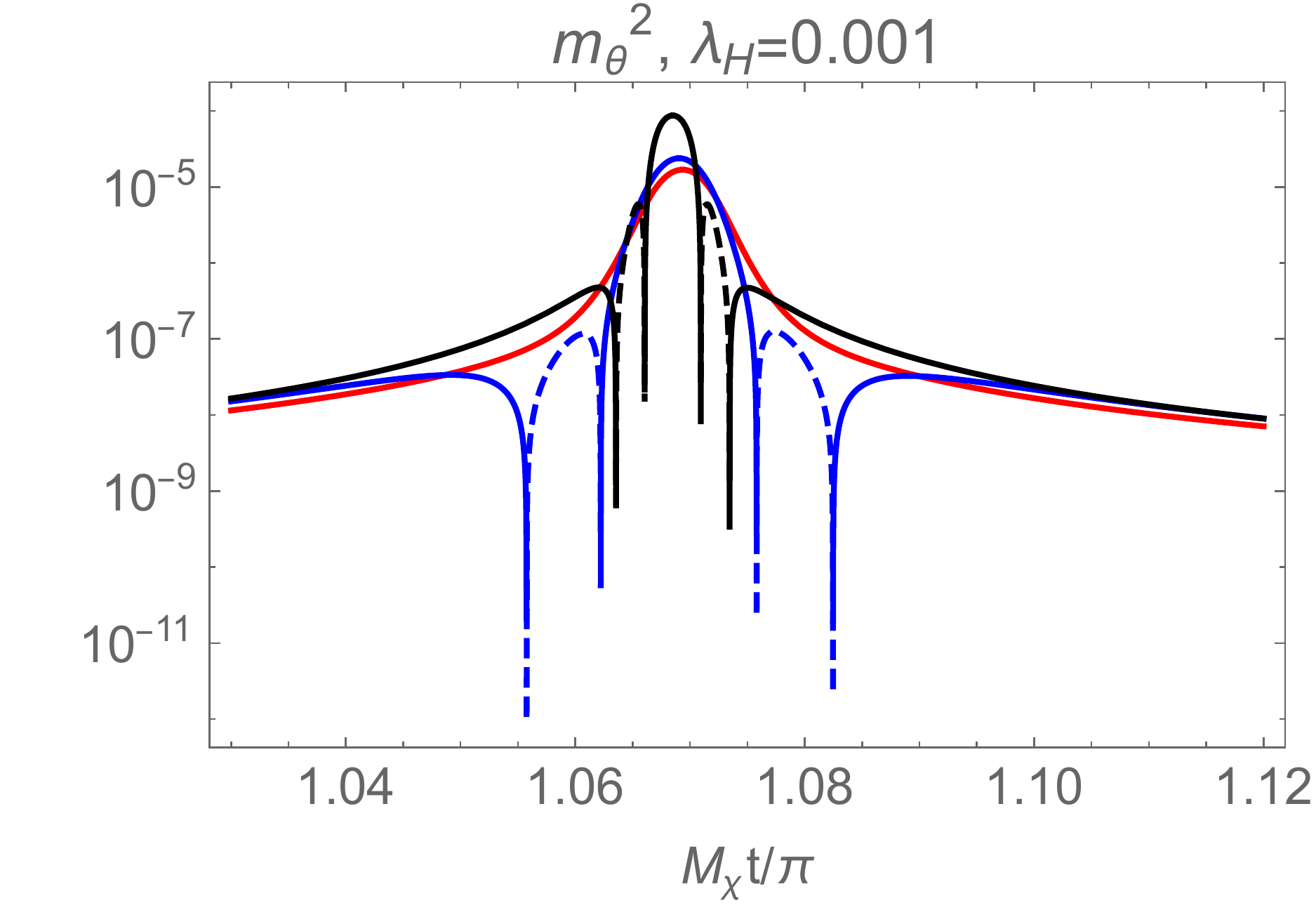}
\includegraphics[width=8cm]{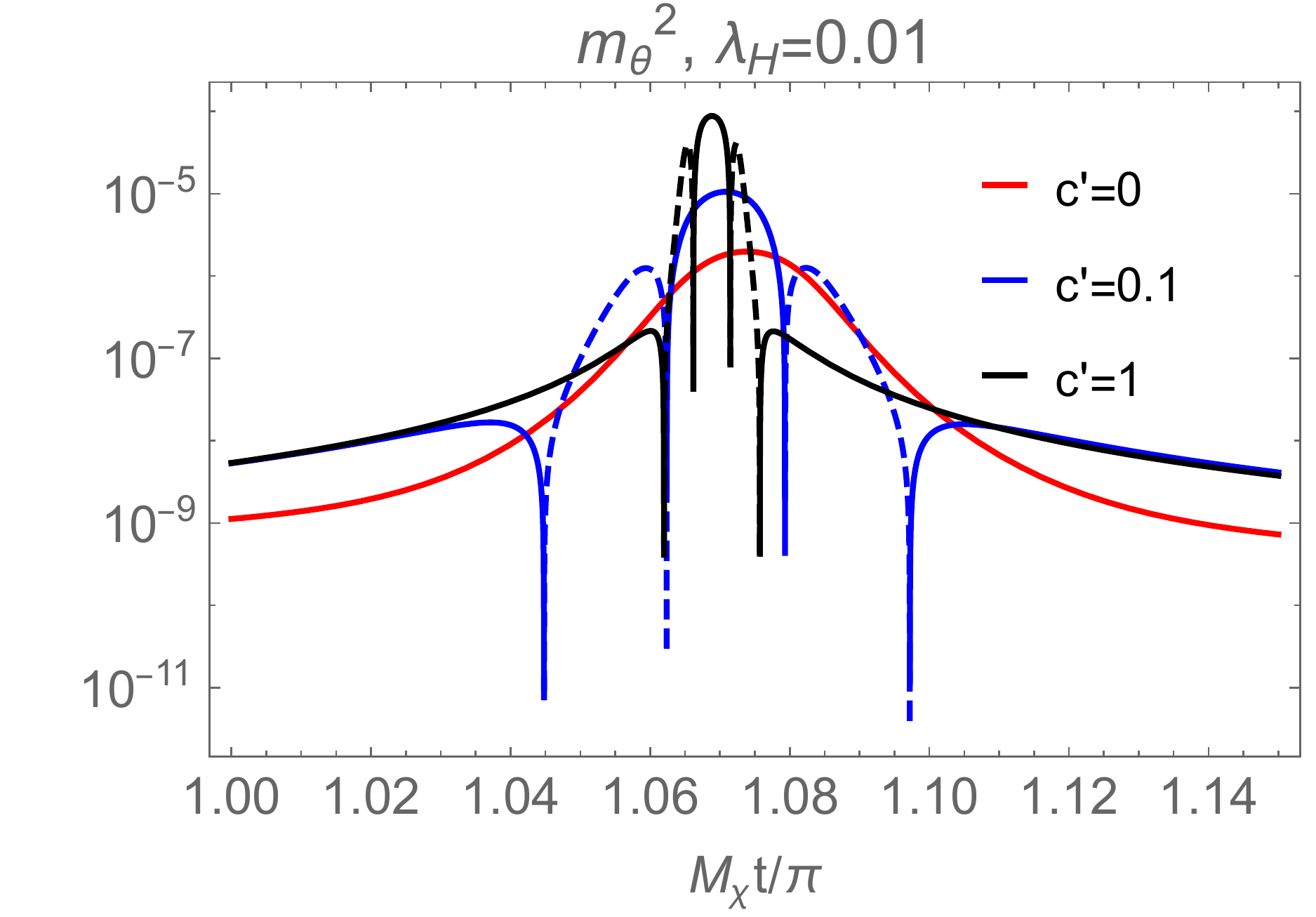}
\caption{
The plots of the effective mass squared of the NG mode. The solid and dashed lines correspond to the positive and negative signs, respectively.
The left and right figures correspond to $\lambda_H=10^{-3}$ and $10^{-2}$, respectively. 
The unit of the time is Eq.~\eqref{Eq:M_chi}.
}
\label{fig:NGmass}
\end{center}
\end{figure}
\begin{figure}[t]
\begin{center}
\includegraphics[width=8cm]{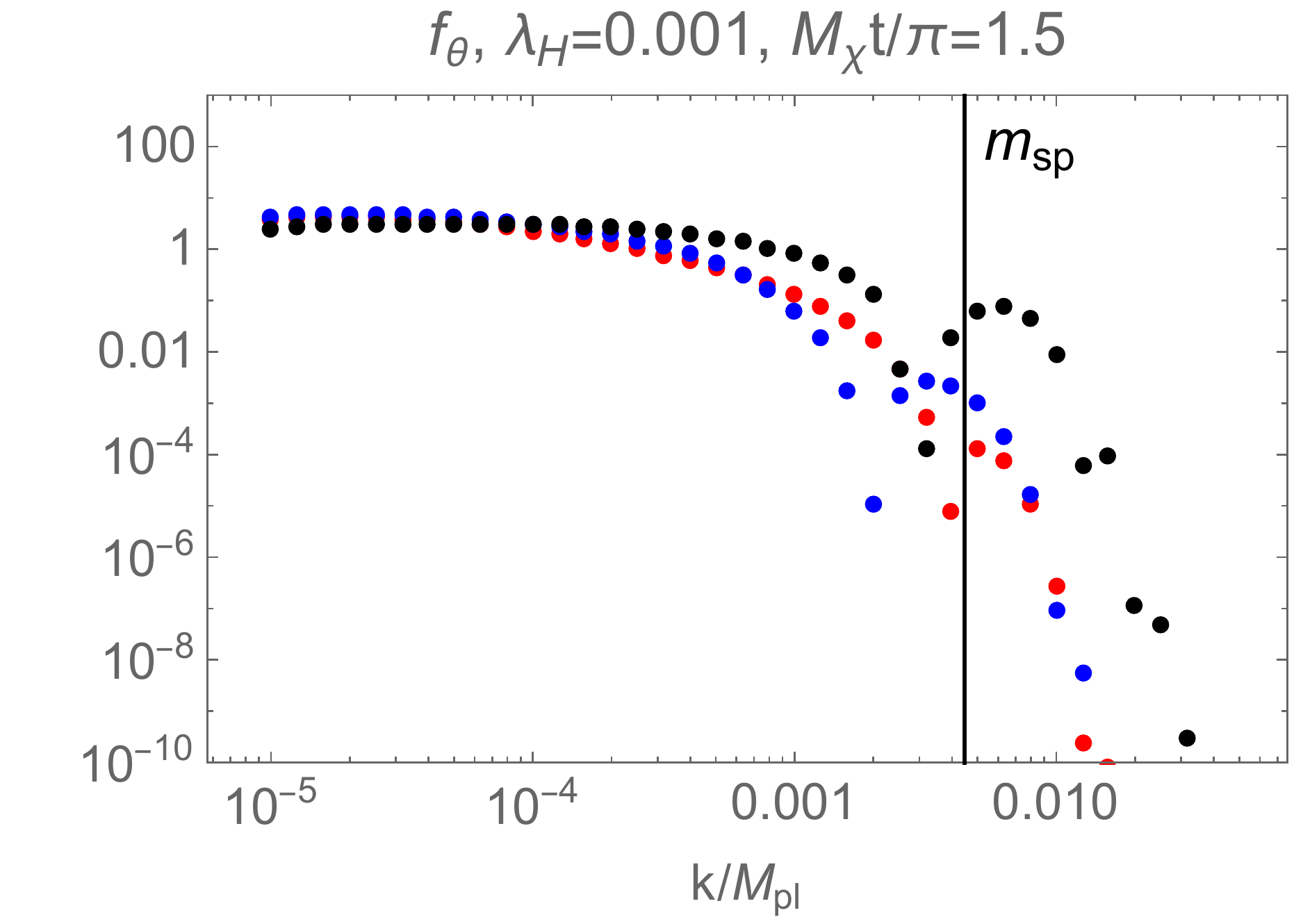}
\includegraphics[width=8cm]{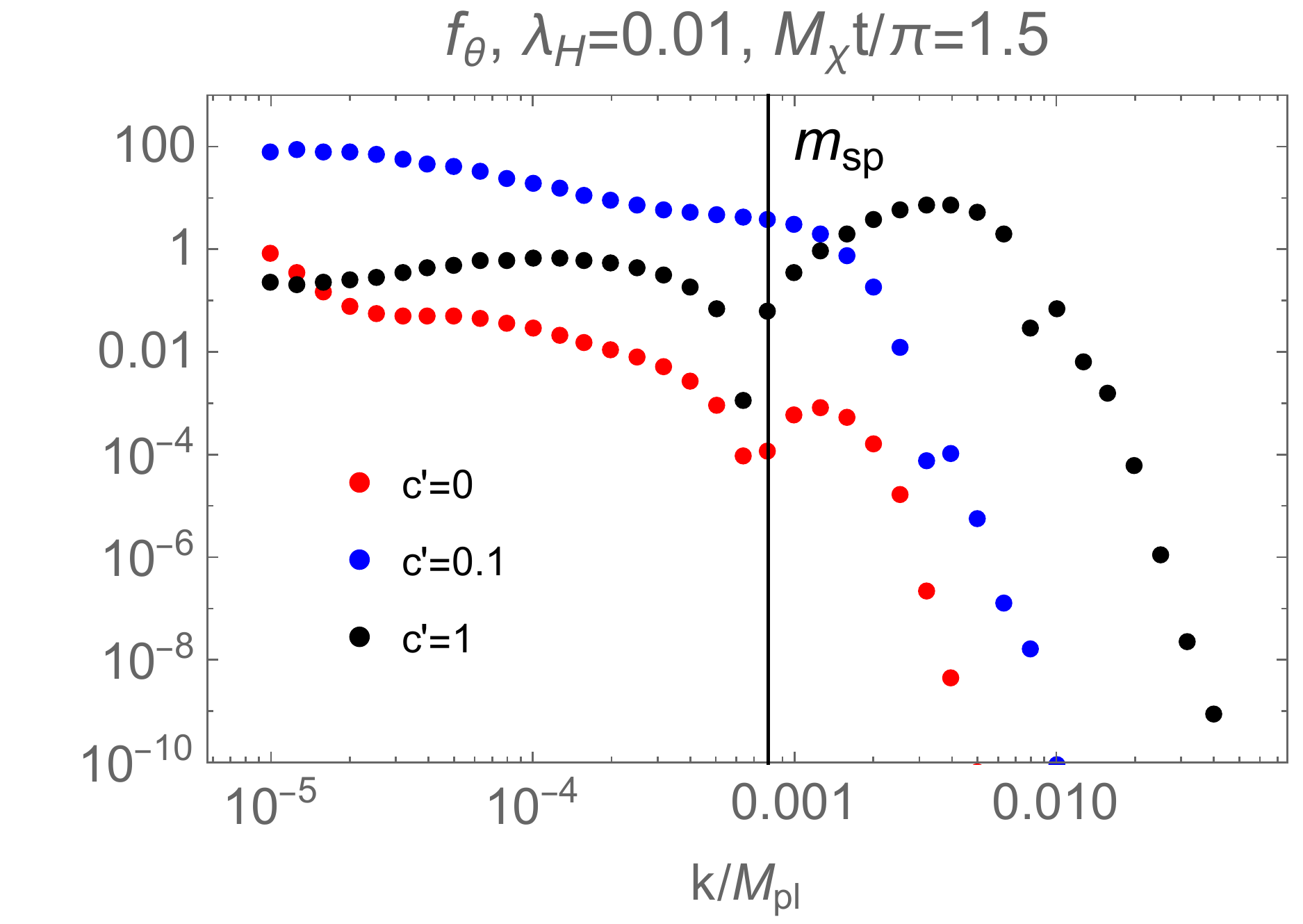}
\caption{
The plots of the occupation number of the NG mode $\theta$, $f_\theta$, after the first zero-crossing ($t=1.5\pi/M_\chi$).
The left and right figures correspond to $\lambda_H=10^{-3}$ and $10^{-2}$, respectively.
The vertical line is the scale given in Eq.~(\ref{eq:new time scale}). 
}
\label{fig:NGspec}
\end{center}
\end{figure}
\begin{figure}[t]
\begin{center}
\includegraphics[width=8cm]{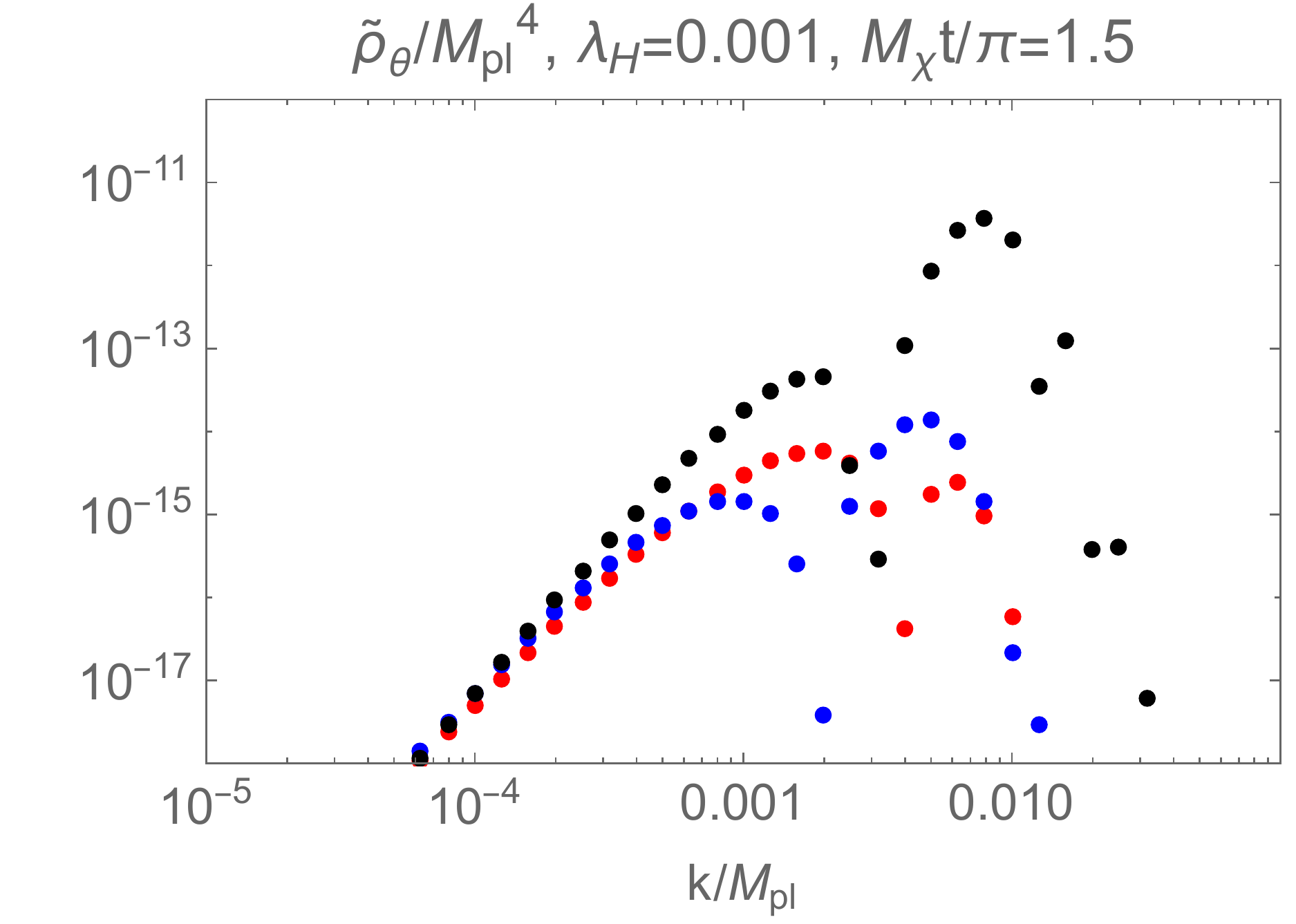}
\includegraphics[width=8cm]{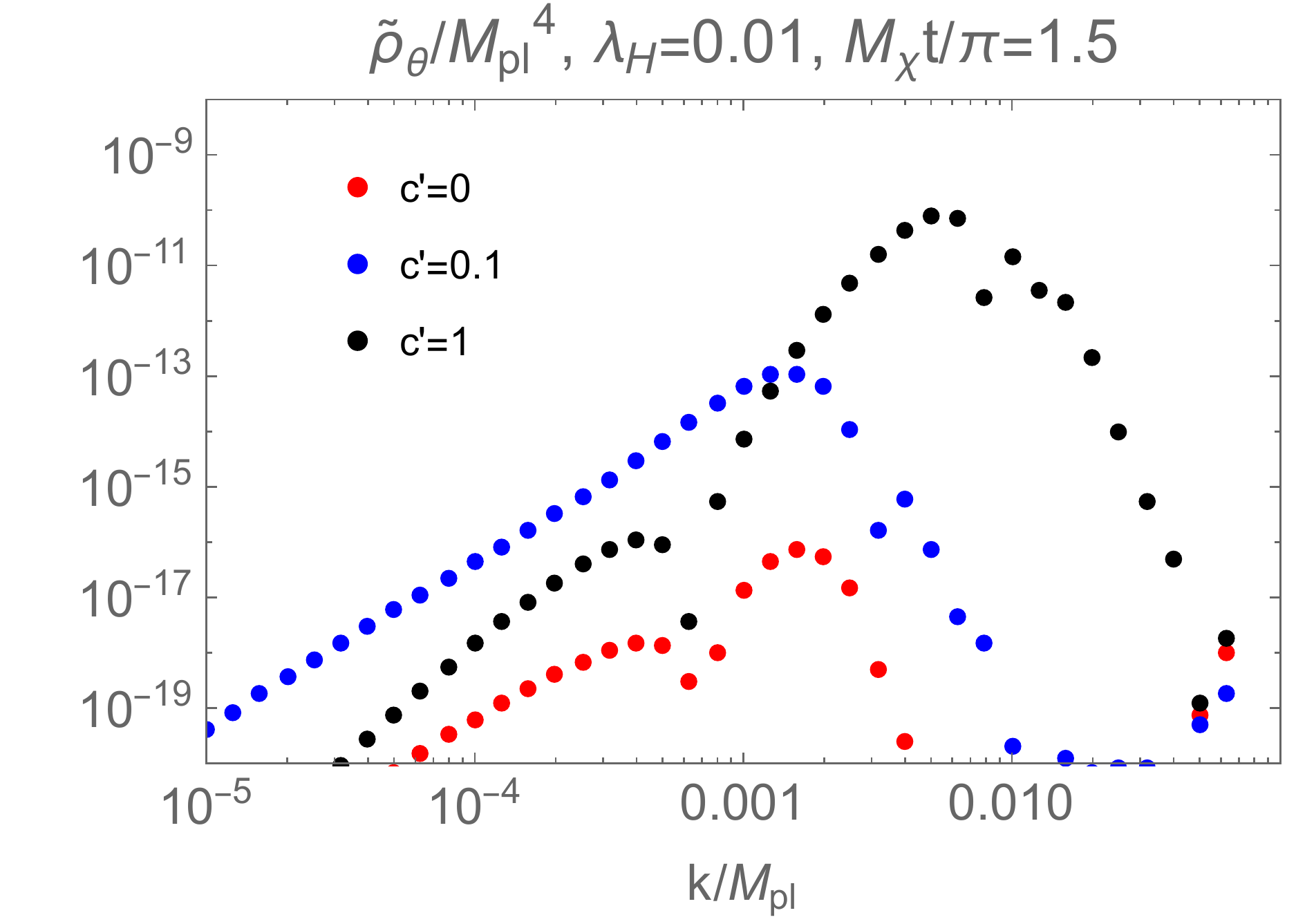}
\caption{
The plots of the energy density spectrum of the NG mode, $\tilde{\rho}_\theta$, after the first zero-crossing ($t=1.5\pi/M_\chi$).
The left and right figures correspond to $\lambda_H=10^{-3}$ and $10^{-2}$, respectively. 
}
\label{fig:NGspec_2}
\end{center}
\end{figure}
\begin{figure}[t]
\begin{center}
\includegraphics[width=8cm]{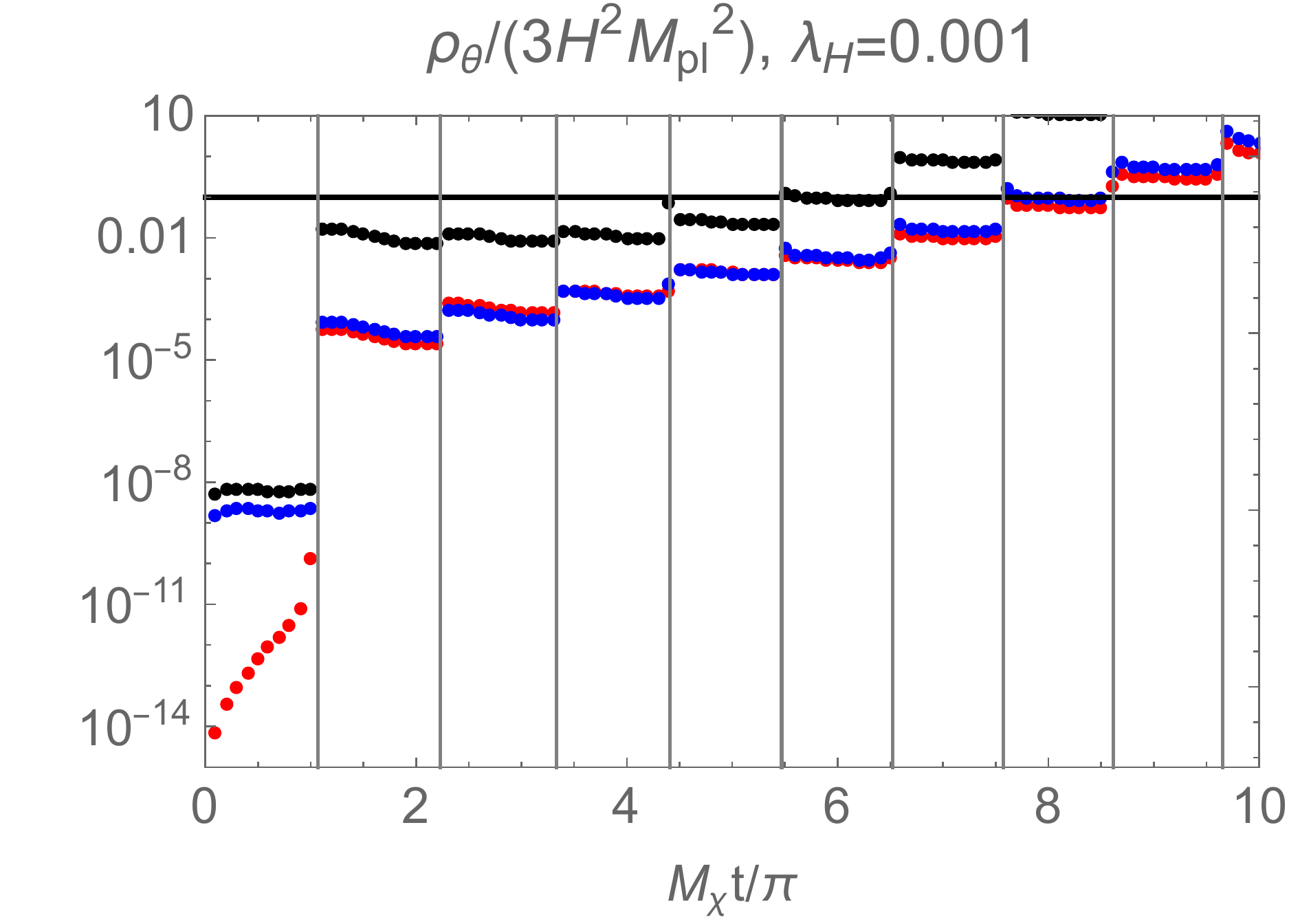}
\includegraphics[width=8cm]{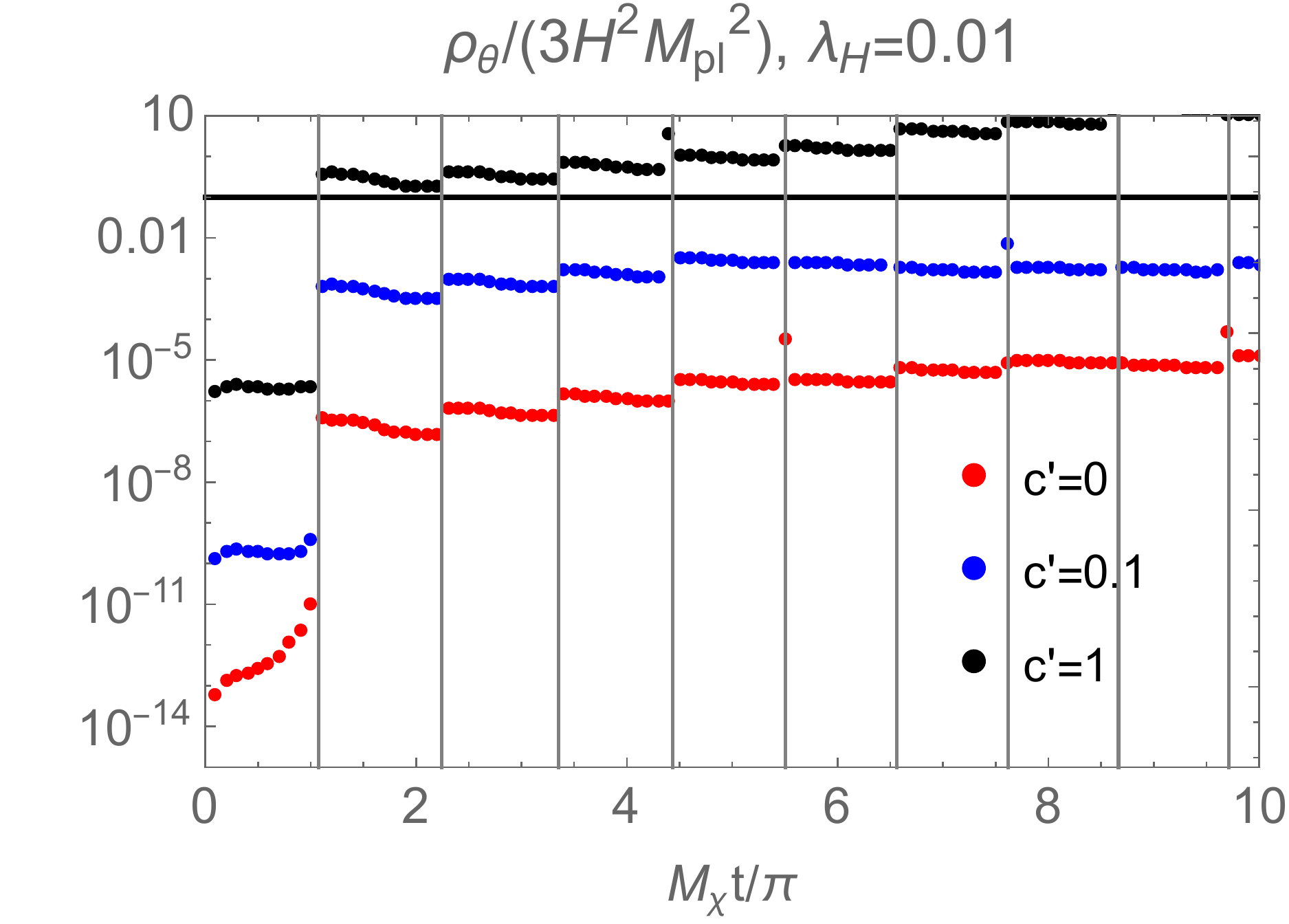}
\caption{
The plots of the energy density, $\rho_\theta$, normalized by the background energy density as functions of $t$. The gray vertical lines correspond to zero-crossings. The black horizon line corresponds to the value $0.1$, where the backreaction may become important. The left and right figures correspond to $\lambda_H=10^{-3}$ and $10^{-2}$, respectively.  The unit of the time is Eq.~\eqref{Eq:M_chi}.
}
\label{fig:NGspec_3}
\end{center}
\end{figure}

In Fig.~\ref{fig:NGmass}, we plot the effective mass of the NG mode around the first zero-crossing. The solid and dashed lines correspond to the positive and negative signs, respectively. We observe that, for $c'=0.1$ and $1$, the mass square can be tachyonic. Moreover, the maximum value of $m_\theta^2$ is larger than $c'=0$ due to the contribution in Eq.~\eqref{Eq:new_term}. 
For these reasons, we expect that the particle production for $c'\neq0$ is more efficient than $c'=0$ case.

In Figs.~\ref{fig:NGspec} and \ref{fig:NGspec_2}, we show the numerical calculation of $f_\theta^{}(k)$ and $\tilde{\rho}_\theta^{}(k)$ after the first zero-crossing of $\phi_1^{}$ ($t=1.5\pi/M_\chi$).   
The value of the cut off (damping) scale is consistent with our estimate Eq.~(\ref{eq:new time scale}) within an order of magnitude. The left and right figures correspond to $\lambda_H=10^{-3}$ and $10^{-2}$, respectively.

As for the total energy density $\rho_\theta$, the numerical calculation is shown in Fig.~\ref{fig:NGspec_3}. The vertical axis is normalized by the background energy density $3H^2M_{pl}^2$. The gray vertical lines correspond to zero-crossings. The black horizon line corresponds to the value $0.1$, where the backreaction may become important. 
We find that, for $\lambda_H=10^{-3}$, the backreaction becomes important within ten zero-crossings.
On the other hand, for $\lambda_H=10^{-2}$ and $c'=0,\  0.1$, the energy density $\rho_\theta^{}$ is smaller than the background energy density for $t\leq 10\pi/M_\chi$. In these cases, we observe that $\rho_\theta$ is fitted by
\aln{
\frac{\rho_{\theta}}{3H^2M_{pl}^2}\simeq
\begin{cases}
10^{-7} e^{0.5t M_\chi/\pi}\quad\text{for $(\lambda_H,c')=(10^{-2},0)$}\\
2\times10^{-4} e^{0.3t M_\chi/\pi}\quad\text{for $(\lambda_H,c')=(10^{-2},0.1)$}
\end{cases}.
}
If we can extrapolate this expression to larger $t$, we find that $\rho_{\theta}/(3H^2M_{pl}^2)$ reaches $0.1$ at the time
\aln{
\frac{t M_\chi}{\pi}\simeq\begin{cases}
27\quad\text{for $(\lambda_H,c')=(10^{-2},0)$}\\
21\quad\text{for $(\lambda_H,c')=(10^{-2},0.1)$}
\end{cases}.
}

\subsection{$U(1)$ Gauge Boson}
In this subsection, we consider the Abelian Higgs model:  
\aln{
S&=\int d^4x\sqrt{-g_J^{}}\left(
-\frac{1}{4}g_J^{\mu\alpha}g_J^{\nu\beta}F_{\mu\nu}^{}F_{\alpha\beta}^{}-g_J^{\mu\nu}(D_\mu^{}\phi_J^{})^\dagger(D_\nu^{}\phi_J^{})-V_J^{}(|\phi_J^{}|)+\frac{c}{\Lambda_J^4}\left[g_J^{\mu\nu}(D_\mu^{}\phi_J^{})^\dagger(D_\nu^{}\phi_J^{})\right]^2
\right).
}
By moving to the Einstein frame and focusing on the kinetic part of the gauge boson, we have (See also Refs.~\cite{Lozanov:2016pac,Ema:2016dny})
\aln{
S_A^{}&:=\int d^4x\sqrt{-g_E^{}}
\left(-\frac{1}{4}g_E^{\mu\alpha}g_E^{\nu\beta}F_{\mu\nu}^{}F_{\alpha\beta}^{}-\frac{g^2}{2}r_J^2\left(\frac{1}{\Omega^2}+\frac{c\dot{r}_J^2}{\Lambda_J^4}\right)g_E^{\mu\nu}A_\mu^{}A_\nu^{}
\right)
\nn
&=\int d^4x a(t)^3
\left(\frac{1}{2a(t)^2}(\dot{A}_i^{}-\partial_i^{}A_0^{})^2-\frac{1}{4a^4}\sum_{i,j=1}^3(\partial_i^{}A_j^{}-\partial_j^{}A_i^{})^2
-\frac{g^2}{2}r_J^2\left(\frac{1}{\Omega^2}+\frac{c\dot{r}_J^2}{\Lambda_J^4}\right)(-A_0^{2}+a(t)^{-2}\mathbf{A}^2)
\right)
\nn
&=\int d\eta d^3x 
\left(\frac{1}{2}(A_i^{'}-\partial_i^{}A_\eta^{})^2-\frac{1}{2}|\nabla\times \mathbf{A}|^2
-\frac{g^2a^2}{2}r_J^2\left(\frac{1}{\Omega^2}+\frac{c\dot{r}_J^2}{\Lambda_J^4}\right)(-A_\eta^{2}+\mathbf{A}^2)
\right)
\nn
&=\frac{1}{2}\int d\eta \int \frac{d^3k}{(2\pi)^3} 
\left((k^2+m_A^2)|\tilde{A}_\eta^{}|^2-i\mathbf{k}\cdot \tilde{\mathbf{A}}'\tilde{A}_\eta^{}+(\text{h.c.})+\tilde{\mathbf{A}}^{'2}-|\mathbf{k}\times \tilde{\mathbf{A}}|^2-m_A^2|\tilde{\mathbf{A}}|^2
\right)
\nn
&=\frac{1}{2}\int d\eta \int \frac{d^3k}{(2\pi)^3} 
\left((k^2+m_A^2)\left|\tilde{A}_\eta^{}-\frac{i\mathbf{k}\cdot \tilde{\mathbf{A}}'}{k^2+m_A^2}\right|^2-\frac{|\mathbf{k}\cdot \tilde{\mathbf{A}}'|^2}{k^2+m_A^2}+|\tilde{\mathbf{A}}'|^2-|\mathbf{k}\times \tilde{\mathbf{A}}|^2-m_A^2|\tilde{\mathbf{A}}|^2
\right), \label{eq:action for gauge boson}
}
where the unitary gauge is taken, $(\tilde{A}_\eta,\tilde{\mathbf{A}})$ represent the Fourier modes, $r_J^{}$ is the radial component of $\phi_J^{}$, $g$ is a $U(1)$ coupling, and  
\aln{
m_A^{}:=\frac{gar_J}{\Omega}\sqrt{1+\frac{c\,\Omega^2\dot{r}_J^2}{\Lambda_J^4}}. 
\label{mass of gauge}
}
After integrating $\tilde{A}_\eta^{}$, we have
\aln{
S_A\to S_A'=\frac{1}{2}\int d\eta \int \frac{d^3k}{(2\pi)^3} 
\left(|\tilde{\mathbf{A}}'|^2-\frac{|\mathbf{k}\cdot \tilde{\mathbf{A}}'|^2}{k^2+m_A^2}-|\mathbf{k}\times \tilde{\mathbf{A}}|^2-m_A^2|\tilde{\mathbf{A}}|^2
\right). 
}
Then, by decomposing $\tilde{\mathbf{A}}$ into the transverse and longitudinal modes as 
\aln{
\tilde{\mathbf{A}}:= \tilde{\mathbf{A}}_T^{}+\frac{\mathbf{k}}{|\mathbf{k}|}\tilde{A}_L^{},\quad \tilde{\mathbf{A}}_T^{}\cdot \mathbf{k}=0,
}
we obtain
\aln{
S_A'=&=\frac{1}{2}\int d\eta \int \frac{d^3k}{(2\pi)^3} 
\left(|\tilde{\mathbf{A}}_T^{'}|^2+|\tilde{A}_L^{'}|^2-\frac{k^2|\tilde{A}_L^{'}|^2}{k^2+m_A^2}-|\mathbf{k}\times \tilde{\mathbf{A}}_T^{}|^2-m_A^2(|\tilde{\mathbf{A}}_T^{}|^2+|\tilde{A}_L^{}|^2)
\right)
\nn
&=\frac{1}{2}\int d\eta \int \frac{d^3k}{(2\pi)^3} 
\left(|\tilde{\mathbf{A}}_T^{'}|^2-(k^2+m_A^2)|\tilde{\mathbf{A}}_T^{}|^2
+\frac{m_A^2|\tilde{A}_L^{'}|^2}{k^2+m_A^2}-m_A^2|\tilde{A}_L^{}|^2 
\right),
}
from which we can see that the transverse component $\mathbf{A}_T^{}$ is already canonically normalized and its mass is given by $m_A^{}$.  
Thus, it does not show any spike-like behavior and its particle production is similar to that of the broad resonance~\cite{Bezrukov:2008ut,GarciaBellido:2008ab}.  

\begin{figure}[t]
\begin{center}
\includegraphics[width=8cm]{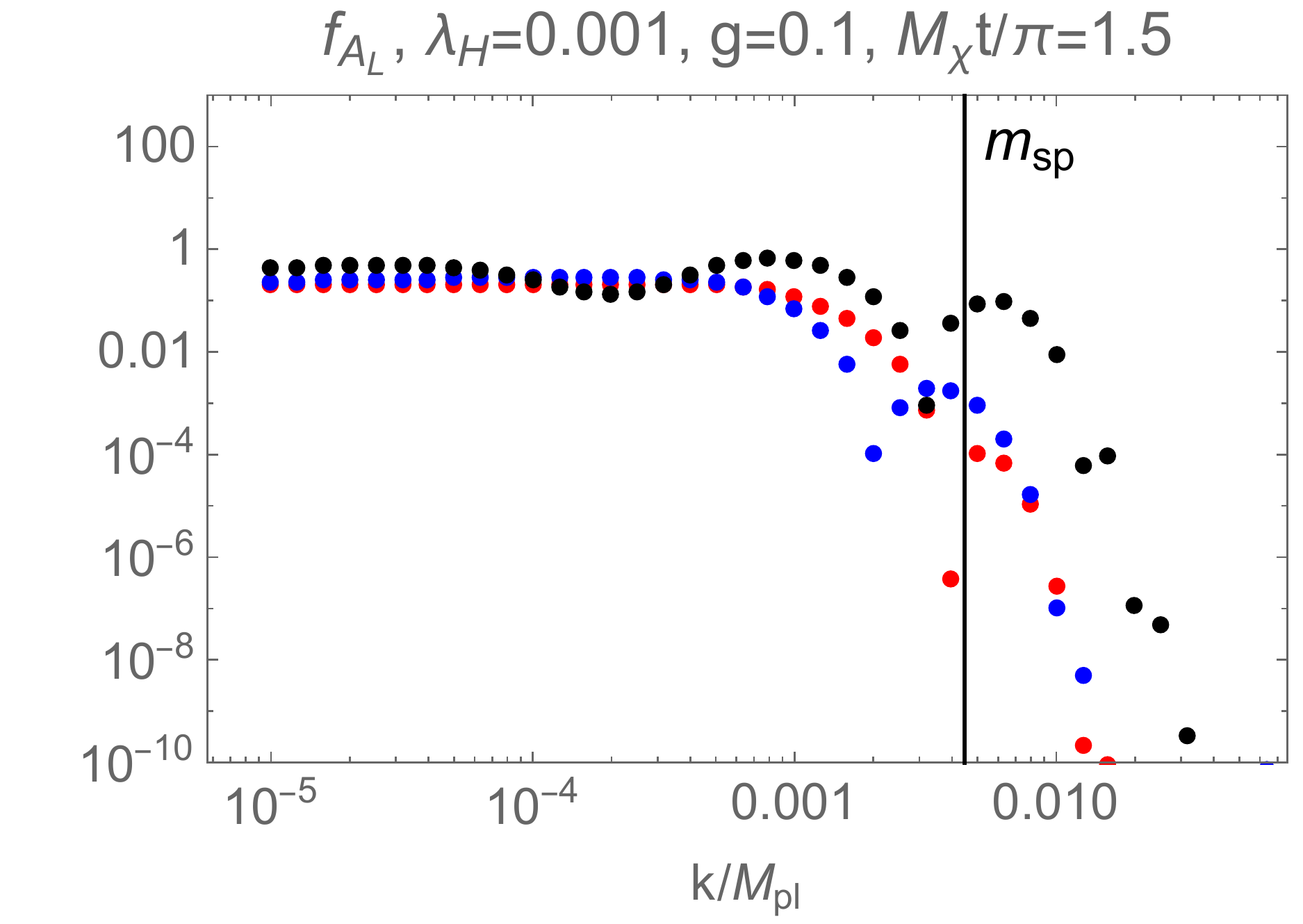}
\includegraphics[width=8cm]{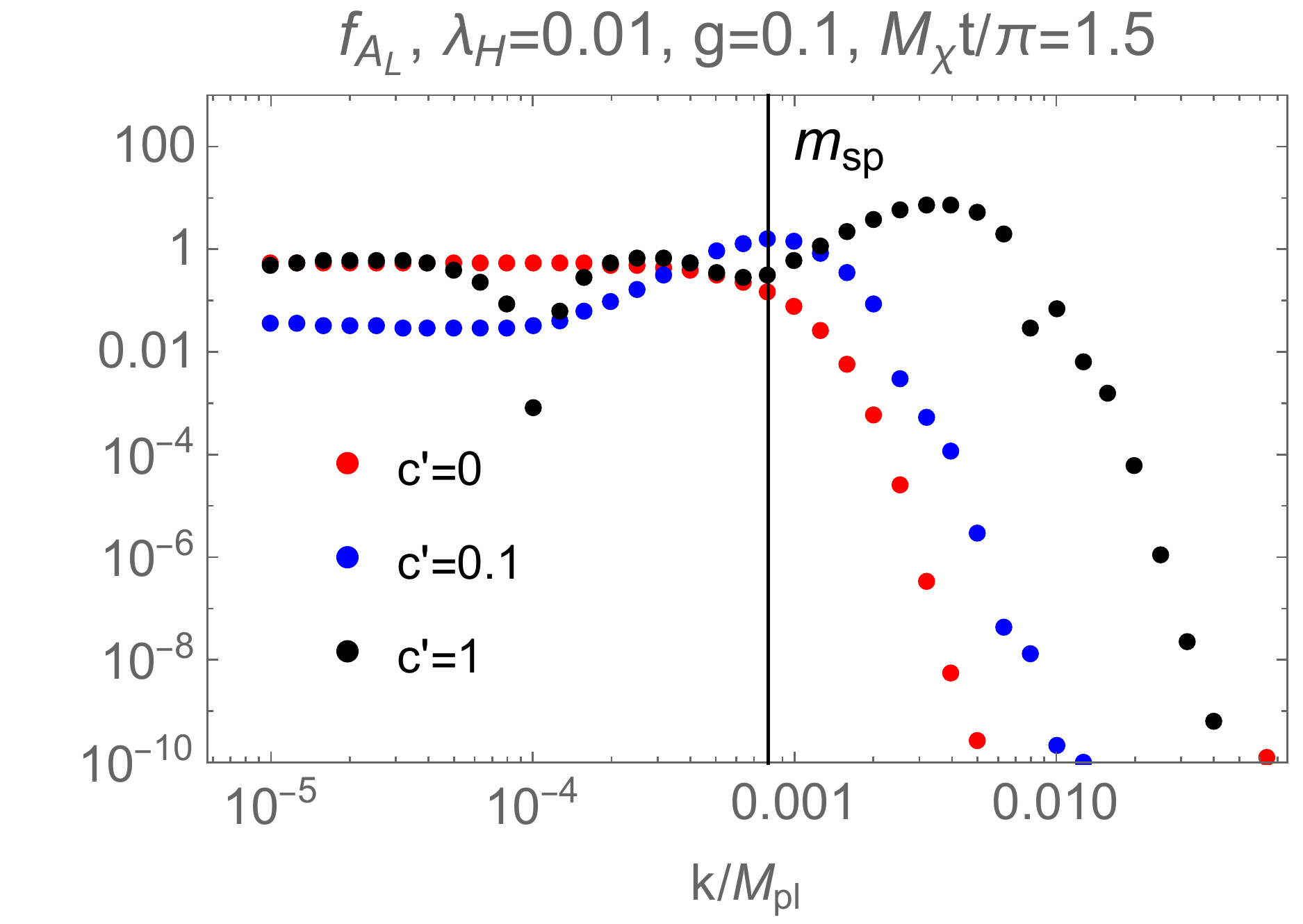}
\caption{
The plots of the occupation number of the longitudinal gauge boson, $f_{A_L}$, after the first zero-crossing ($t=1.5\pi/M_\chi$).
The left and right figures correspond to $\lambda_H=10^{-3}$ and $10^{-2}$, respectively.
The $U(1)$ gauge coupling is taken to be $g=0.1$. The vertical line is the scale given in Eq.~(\ref{eq:new time scale}). 
}
\label{fig:AL}
\end{center}
\end{figure}

\begin{figure}[t]
\begin{center}
\includegraphics[width=8cm]{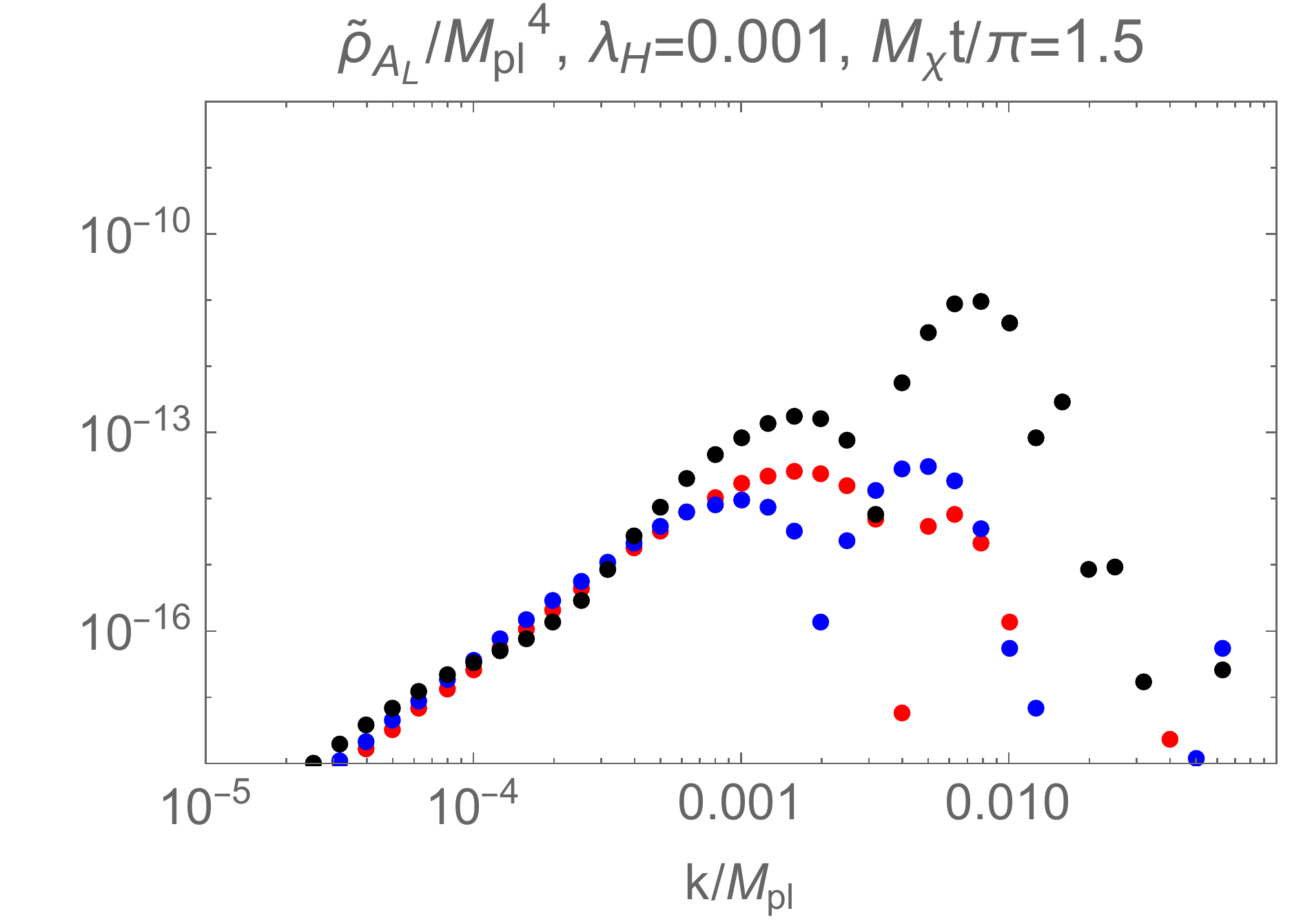}
\includegraphics[width=8cm]{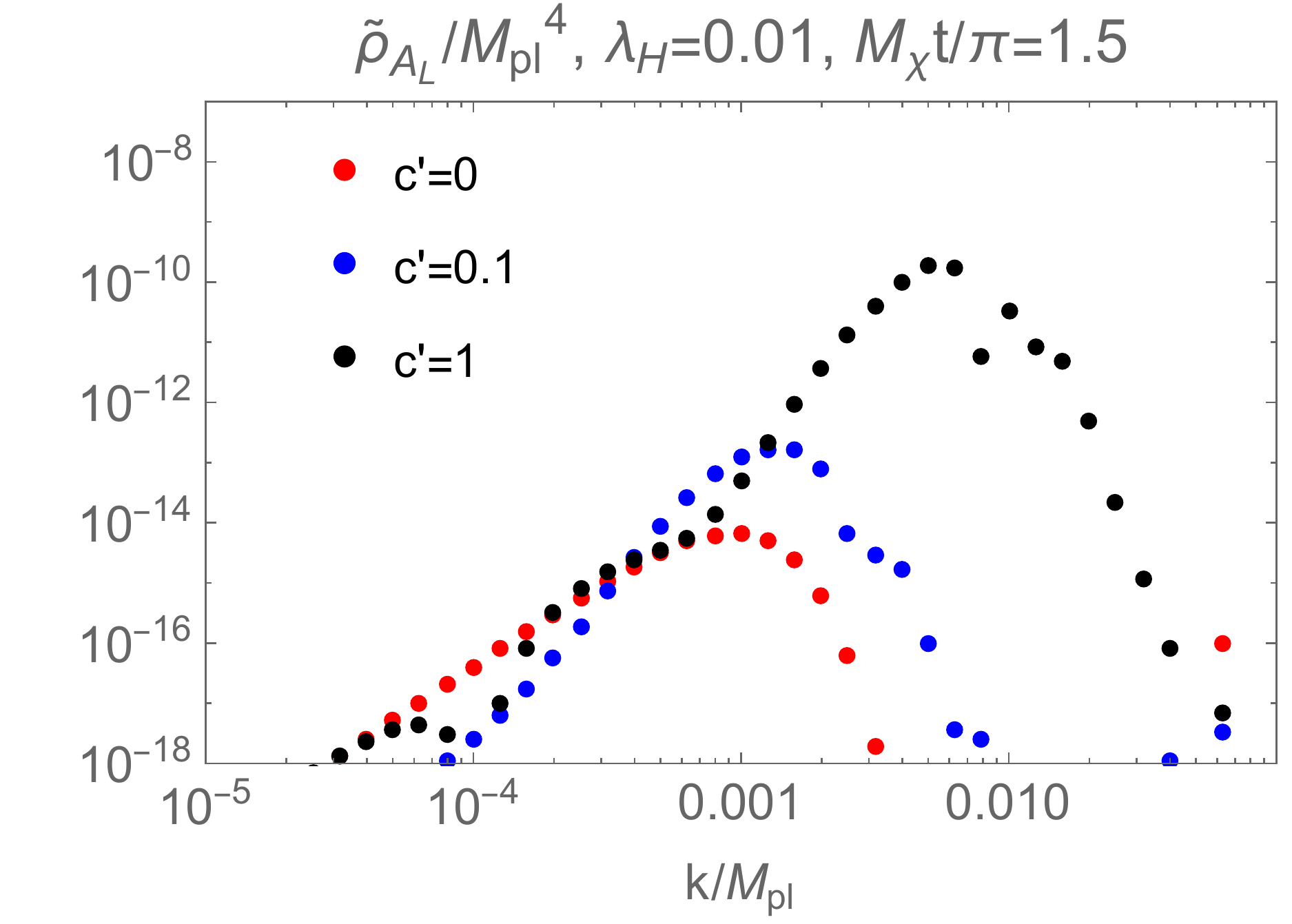}
\caption{
The plots of the energy density spectrum of the longitudinal gauge boson, $\tilde{\rho}_{A_L}$, after the first zero-crossing ($t=1.5\pi/M_\chi$).
The left and right figures correspond to $\lambda_H=10^{-3}$ and $10^{-2}$, respectively.
The $U(1)$ gauge coupling is taken to be $g=0.1$.
}
\label{fig:AL_2}
\end{center}
\end{figure}

On the other hand, the longitudinal component $A_L^{}$ is not canonically normalized, so we have to introduce a new field by  
\aln{
\tilde{{\cal{A}}}_L^{}:= \frac{m_A^{}}{\sqrt{k^2+m_A^2}}\tilde{A}_L^{}. 
}
Then, the action of ${\cal{A}}_L^{}$ becomes 
\aln{
&\frac{1}{2}\int d\eta \int \frac{d^3k}{(2\pi)^3} \left(\left|\tilde{{\cal{A}}}_L^{'}-\frac{m_A^{'}}{m_A^{}}\frac{k^2}{k^2+m_A^2}\tilde{{\cal{A}}}_L^{}\right|^2-(k^2+m_A^2)|\tilde{{\cal{A}}}_L^{}|^2 \right)
\nn
=&\frac{1}{2}\int d\eta \int \frac{d^3k}{(2\pi)^3} \left[\left|\tilde{{\cal{A}}}_L^{'}\right|^2-\left\{k^2+m_A^2-\frac{d}{d\eta}\left(\frac{m_A^{'}}{m_A^{}}\frac{k^2}{k^2+m_A^2}\right)-\left(\frac{m_A^{'}}{m_A^{}}\frac{k^2}{k^2+m_A^2}\right)^2\right\}|\tilde{{\cal{A}}}_L^{}|^2 \right].
}
The particle production is computed by using Eq.~\eqref{Eq:fluctuations} or \eqref{Eq:Bogiliubov} with
\aln{
\omega_k^2(\eta)=k^2+m_A^2-\frac{d}{d\eta}\left(\frac{m_A^{'}}{m_A^{}}\frac{k^2}{k^2+m_A^2}\right)-\left(\frac{m_A^{'}}{m_A^{}}\frac{k^2}{k^2+m_A^2}\right)^2.
}
 One can show that the effective mass obtained from this action coincides with that of the NG boson Eq.(\ref{NG effective mass}) for $k\gg m_A^{}$ since $m_A^{}$ corresponds to $F$.
As in the NG boson case, we find that it is possible to become $\omega_k^2<0$ for nonzero $c'$.
As long as we focus on the main production mode $k\sim m_{\rm sp}^{}$, the mass difference is characterized by a small parameter 
\aln{
\varepsilon
:&=\left(1-\frac{k^2}{k^2+m_A^2}\right)\bigg|_{k=m_{\rm sp}^{}}^{}
\\
&\sim 10^{-3} \times c^{1/2}\left(\frac{g}{0.1}\right)^2\left(\frac{\lambda_H^{}}{10^{-3}}\right)^{2} \quad \text{for }r_J^{}\lesssim \tilde{\phi}=\left(\lambda_H^{}/\xi\right)^{1/2}M_{pl}^{},
}
where we have used Eqs. (\ref{eq:lambdaH_xi})(\ref{phitilde}). 
Thus, we expect that the qualitative behavior of particle production is similar between the NG boson and the longitudinal gauge boson for $k\gtrsim m_A^{}$ except for the small corrections characterized by $\varepsilon$.
In Figs.~\ref{fig:AL} and \ref{fig:AL_2}, we show our numerical calculation of $f_{A_L^{}}^{}(k)$ and $\tilde{\rho}_{A_L^{}}^{}(k)$ after the first zero-crossing of $\phi_1^{}$.  
Here, we choose $g=0.1$.
As in the NG boson case, we see that the damping scale is roughly given by $m_\text{sp}$ in Eq.~\eqref{eq:new time scale}.

\begin{figure}[t]
\begin{center}
\includegraphics[width=8cm]{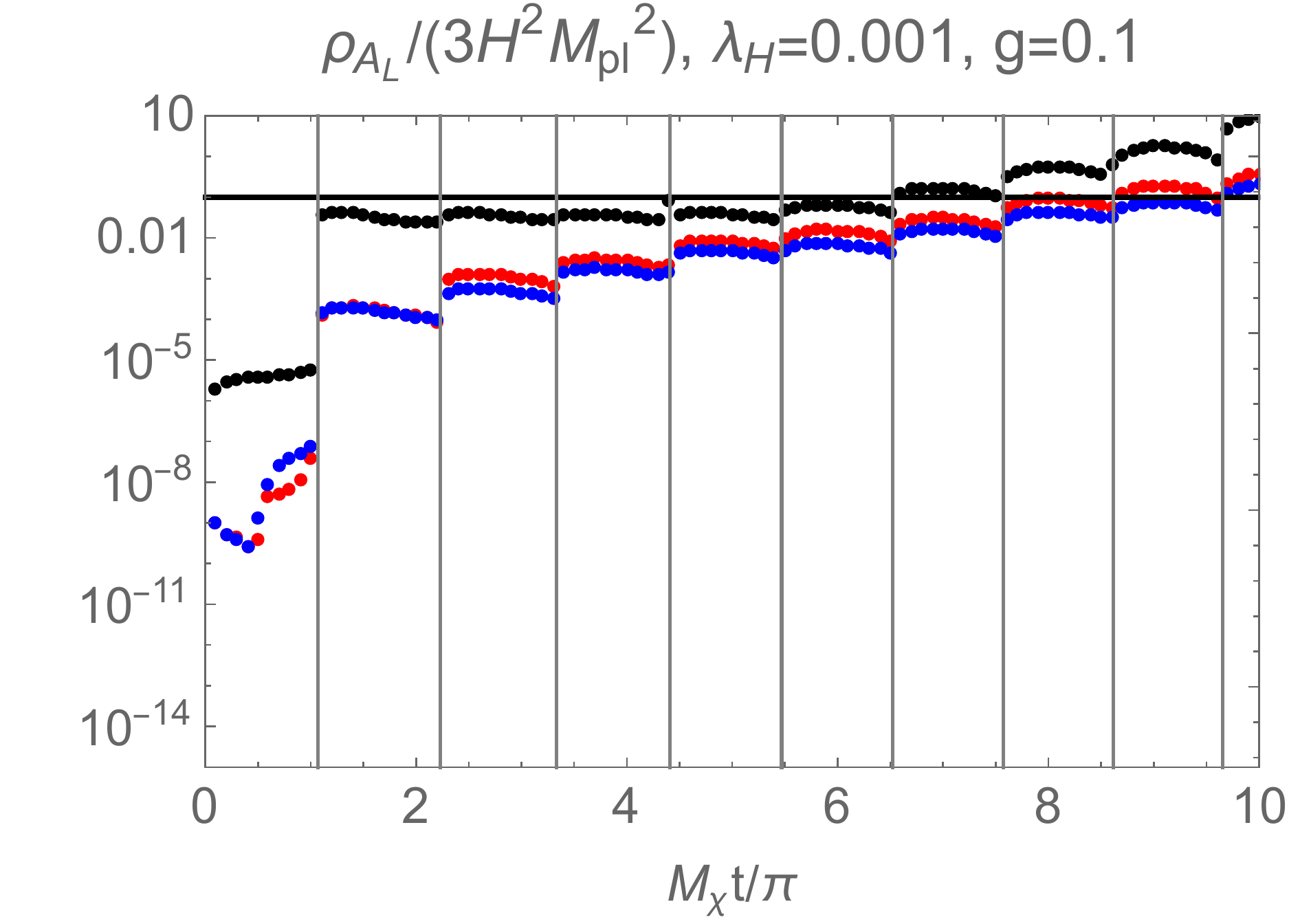}
\includegraphics[width=8cm]{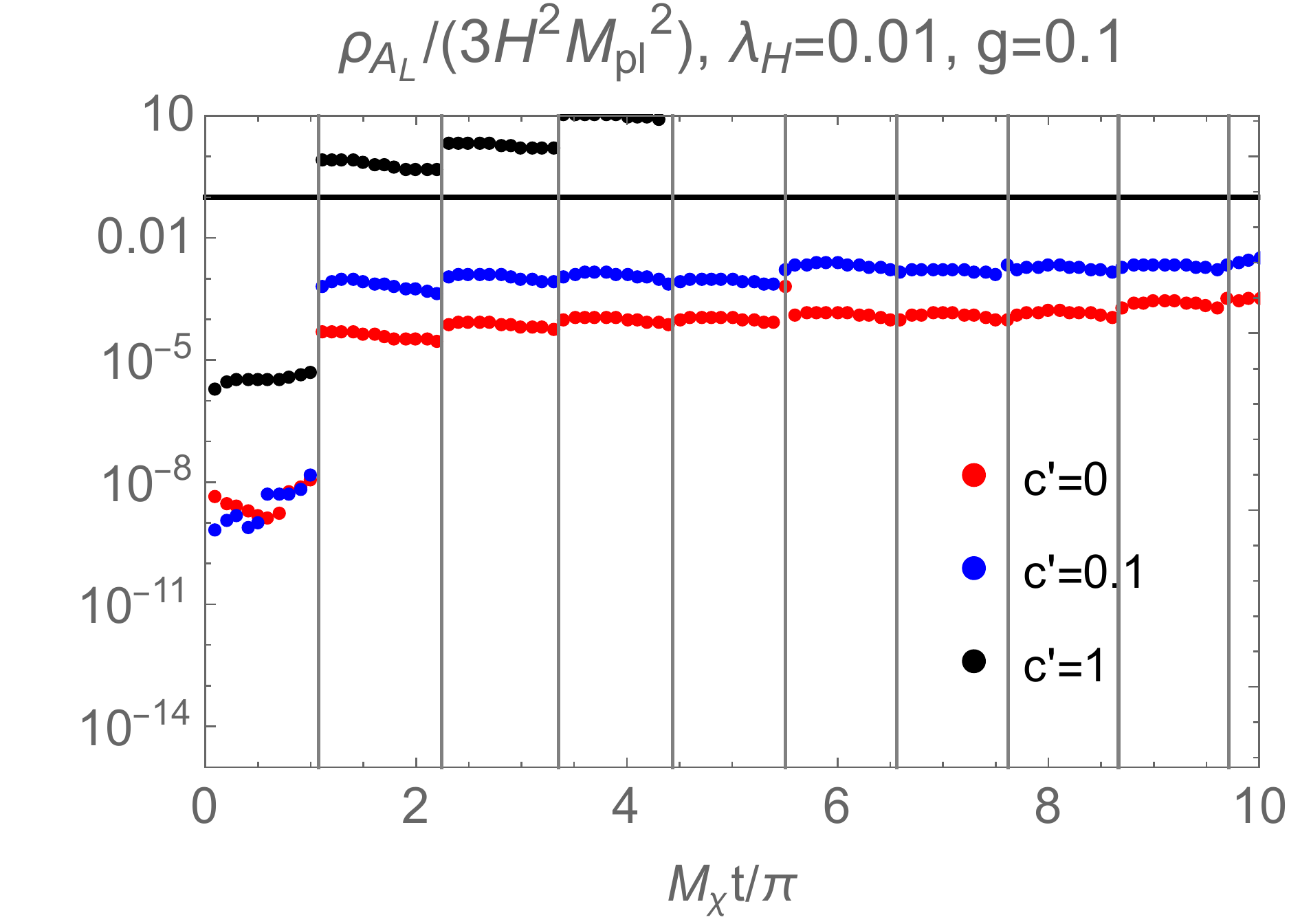}
\caption{
The plots of the energy density, $\rho_{A_L}$, normalized by the background energy density as functions of $t$. The gray vertical lines correspond to zero-crossings. The black horizon line corresponds to the value $0.1$, where the backreaction may become important. The left and right figures correspond to $\lambda_H=10^{-3}$ and $10^{-2}$, respectively. The unit of the time is Eq.~\eqref{Eq:M_chi}.
}
\label{fig:AL_3}
\end{center}
\end{figure}

In Fig.~\ref{fig:AL_3}, the total energy densities of the longitudinal gauge boson are plotted
as functions of $t$. As in Fig.~\ref{fig:NGspec_3}, the vertical axis is normalized by the background energy density $3H^2M_{pl}^2$. The gray vertical lines correspond to zero-crossings, and the black horizon line corresponds to the value $0.1$.
We observe that the backreaction is important within the first ten zero-crossings except for $(\lambda_H,c')=(10^{-2},0)$ and $(10^{-2},0.1)$. In these cases, the energy densities are fitted by
\aln{
\frac{\rho_{A_L}}{3H^2M_{pl}^2}\simeq
\begin{cases}
4\times10^{-5} e^{0.2t M_\chi/\pi}\quad\text{for $(\lambda_H,c')=(10^{-2},0)$}\\
5\times10^{-4} e^{0.2t M_\chi/\pi}\quad\text{for $(\lambda_H,c')=(10^{-2},0.1)$}
\end{cases}.
} 
If we can extrapolate this expression to a larger $t$, we find that the backreaction becomes important at the time
\aln{
\frac{t M_\chi}{\pi}\simeq\begin{cases}
39\quad\text{for $(\lambda_H,c')=(10^{-2},0)$}\\
26\quad\text{for $(\lambda_H,c')=(10^{-2},0.1)$}
\end{cases}.
}

In a realistic situation, we notice that the longitudinal gauge boson decays into the light fermions in the SM. The energy density is expected to be smaller when we take into account the effect of the decay.


\section{Summary}\label{sec:Summary}
We have argued that the physics of Higgs inflation during the era of preheating is highly sensitive to the presence of higher-dimensional operators, and is therefore UV-dependent. 
In particular, we have shown that the operator $|\partial_\mu H|^4$ can dramatically alter the behavior of the Higgs dynamics during preheating and suppress the violent spike-like behavior that otherwise leads to high-energy particle production. 
This suppression maintains the consistency of the effective theory and avoids the unitarity problems associated with Higgs inflation. 

As discussed above, there are many other higher-dimensional operators that can affect preheating. Further research is needed to determine which conditions and UV models allow unitarity to be preserved during Higgs inflation.

In order to have a better understanding of the preheating dynamics, it may be necessary to perform a lattice simulation that takes into account the non-linear effect (see e.g. Ref.~\cite{Nguyen:2019kbm} for a study of the preheating for the multifield inflation with non-minimal couplings).
It is interesting to perform the lattice simulation with the higher dimensional operators. 

Higgs inflation is a promising avenue for obtaining inflationary physics within the Standard Model. The UV-sensitivity of the unitarity problem allows us to constrain possible UV completions of this model, offering a valuable window into even higher-energy physics.

\section*{Acknowledgements} 
The work of YH is supported by the Advanced ERC grant SM-grav, No 669288.
The work of KK is supported by the Grant-in-Aid for JSPS Research Fellow, Grant Number 17J03848. 
The work of AS is supported by DOE Grant DE-SC0012012, NSF Grant PHY-1720397, the Heising-Simons Foundation Grants 2015-037, and the Gordon and Betty Moore Foundation Grant GBMF7946.
YH thanks the hospitality of the Kavli Institute for Theoretical Physics (supported by NSF PHY-1748958) where part of this work was carried out.

\appendix
\def\thesection{Appendix \Alph{section}}
\section{Unitarity Issue in Higgs Inflation}\label{app:unitarity}
In the Higgs inflation scenario with large non-minimal coupling $\xi\gg 1$, the naive tree-level cutoff scale $\Lambda=M_{pl}^{}/\xi$ is comparable to or smaller than typical energy scales during inflation, e.g. ${\cal{H}}\sim \Lambda,\ \chi\sim M _{pl}^{}$. 
However, in \cite{Bezrukov:2010jz}, it was argued that the cutoff scale $\Lambda$ is actually a background-dependent quantity and  remains sufficiently large compared with these relevant dynamical scales throughout the whole history of the universe. 
Here, we denote the Higgs in the Jordan frame as $\varphi$ and choose the unitary gauge. 

In the Jordan frame, by expanding the Higgs field and the metric around the backgrounds as $\varphi=\overline{\varphi}+\delta \varphi,\ g_{\mu\nu}^{}=\overline{g}_{\mu\nu}^{}+ h_{\mu\nu}^{}$ and redefining the canonically normalized fields $(\delta \hat{\varphi},\hat{h}_{\mu\nu}^{})$, we obtain various higher-dimensional operators. 
Among them, the leading one is the cubic Higgs-graviton interaction 
\aln{
\frac{\xi\sqrt{M_{pl}^2+\xi\overline{\varphi}^2}}{M_{pl}^2+(\xi+6\xi^2)\overline{\varphi}^2}\delta \hat{\varphi}^2\Box \hat{h}:=\frac{1}{\Lambda_J^{}(\overline{\varphi})}\delta \hat{\varphi}^2\Box \hat{h}
}
where $\hat{h}=\overline{g}^{\mu\nu}\hat{h}_{\mu\nu}^{}$ and $\Lambda_J^{}(\overline{\varphi})$ is the cutoff scale in the Jordan frame. 
This behaves as  
\begin{itemize}
\item For $\overline{\varphi}\ll M_{pl}^{}/\xi$,
\aln{\Lambda_J^{}(\overline{\varphi})\simeq \frac{M_{pl}^{}}{\xi}=\Lambda.\label{eq:cutoff 1}
}
\item For $M_{pl}^{}/\xi\ll \overline{\varphi} \ll M_{pl}^{}/\sqrt{\xi}$,
\aln{
\Lambda \lesssim \Lambda_J^{}(\varphi)\simeq \frac{6\xi \varphi_J^2}{M_{pl}^{}}\ll 6M_{pl}^{}.\label{eq:cutoff 2}
}
\item For $ M_{pl}^{}/\sqrt{\xi}\ll \overline{\varphi}$,
\aln{
M_{pl}^{}\lesssim \Lambda_J^{}(\overline{\varphi})\simeq 6\xi \overline{\varphi}. \label{eq:cutoff 3}
}
\end{itemize}
Thus, because Higgs inflation occurs in the third region, $\Lambda_J^{}(\overline{\varphi})$ is larger than the other dynamical scales ${\cal{H}},\chi$ and consistent. 

In the Einstein frame, the higher-dimensional operators only appear through the Higgs potential 
\aln{V(\chi)=\frac{\lambda_H^{}\varphi(\chi)^4}{4\Omega(\varphi(\chi))^4},
}
where $\varphi(\chi)$ is a solution of 
\aln{
\frac{d\chi}{d\varphi}
=\frac{1}{1+\xi \varphi^2/M_{pl}^2}\sqrt{1+\xi(1+6\xi)\frac{\varphi^2}{M_{pl}^2}}. 
}
Then, we can repeat the same argument by expanding $\chi$ as $\chi=\overline{\chi}+\delta\chi$ and reading the cutoff scales from the expansion
\aln{V(\overline{\chi}+\delta\chi)=V(\overline{\chi})+\sum_{n=1}^{}\frac{1}{n!\Lambda_{n}^{n-4}}\delta \chi^n,
}
where
\aln{\Lambda_{n}^{}:=\frac{d^nV(\chi)}{d\chi^n}\bigg|_{\chi=\overline{\chi}}^{-\frac{1}{n-4}}. 
} 
The results are  
\begin{itemize}
\item For $\overline{\chi}\ll M_{pl}^{}/\xi$,
\aln{
\Lambda_n^{}\sim \frac{M_{pl}^{}}{\xi}\lambda_H^{-\frac{1}{n-4}}.
}
\item For $M_{pl}^{}/\xi\ll \overline{\chi} \ll M_{pl}^{}$,
\aln{
\Lambda_n^{}\sim \frac{\xi\overline{\varphi}^2}{M_{pl}^{}}\left(\frac{\xi^6\overline{\varphi}^6}{\lambda_H^{} M_{pl}^6}\right)^{\frac{1}{n-4}}, 
}
\item For $ M_{pl}^{}/\sqrt{\xi}\ll \overline{\chi}$,
\aln{
\Lambda_n^{}\sim M_{pl}^{}. 
}
\end{itemize}
These are consistent with Eqs.~(\ref{eq:cutoff 1})(\ref{eq:cutoff 2})(\ref{eq:cutoff 3}). 
See the paper \cite{Bezrukov:2010jz} for a treatment including the quantum loop corrections. 

\section{Inflaton Dynamics in Conventional Higgs Inflation}\label{app:dynamics in conventional case}
Here we discuss the background dynamics of the Higgs field after Higgs inflation in the conventional case. 
As the inflaton component, we choose $\phi_1$ in Eq.~\eqref{eq:parametrization}. 
The canonically normalized real inflaton field $\chi$ is given by solving
\aln{
\frac{d\chi}{d\phi_1^{}}
=\frac{1}{1+\xi \phi_1^2/M_{pl}^2}\sqrt{1+\xi(1+6\xi)\frac{\phi_1^2}{M_{pl}^2}}.
\label{eq:relation between varphiJ and varphi}
}
Then, its equation of motion is given by
\aln{
\ddot{\chi}+3{\cal H}\dot{\chi}=-\frac{\partial V^{}}{\partial \chi}\simeq -M_\chi^2 \chi,\quad \chi(0)\sim M_{pl}^{},\ \dot{\chi}(0)\sim 0 
\label{eq:eom fo varphi}
}
where $M_\chi^{2}=\lambda_H M_{pl}^2/(3\xi^2) \sim {\cal H}$.   
If we neglect the friction term, the solution of this equation is just an oscillating solution $\chi(t)\sim M_{pl}^{}\sin(M_\chi^{}t)$. 
Therefore, there is no spike-like behavior for $\chi(t)$. 
On the other hand, the background dynamics of $\phi_1^{}$ have impulsive behavior when $\phi_1^{}$ passes the origin because of the change of the dominant 
kinetic term of $\phi_1^{}$ in Eq.~(\ref{eq:action in Einstein frame}).  
In fact, when $\phi_1^{}\gg M_{pl}^{}/(\sqrt{6}\, \xi)$, the term containing $\partial \log\Omega(\phi_1^{})$ dominates, so the Lagrangian effectively becomes
\aln{{\cal{L}}_\chi^{}\simeq 3\xi^2\left(\frac{\phi_1^{}}{M_{pl}^{}}\right)^2(\partial\phi_1^{})^2-\frac{\lambda_H}{4}\phi_1^4
=2\lambda_H^{}\left(\frac{\phi_1^{}}{M_{\chi}^{}}\right)^2\left(\frac{1}{2}(\partial\phi_1^{})^2-\frac{1}{2}\left(\frac{M_\chi^{}}{2}\right)^2\phi_1^{2}\right),
\label{eq:action just after the inflation}
}
from which we can see that 
\aln{
|\dot{\phi}_1^{}|\sim \frac{M_{pl}^{}}{\sqrt{\xi}}\frac{M_\chi^{}}{2}
\label{eq:phijdot}
}
at $\phi_1\simeq M_p/\sqrt{\xi}$.  
After $\phi_1^{}$ drops below $M_{pl}^{}/(\sqrt{6}\xi)$, the usual kinetic term starts to dominate.  
Therefore, from energy conservation, we have
\aln{
&\frac{1}{2}\dot{\phi}_1^2\simeq V_I^{}=\frac{\lambda M_{pl}^4}{4\xi^2}\quad
&&\Rightarrow 
&&\dot{\phi}_1^{}\simeq \sqrt{\frac{3}{2}}M_{pl}^{}M_\chi^{},
\label{eq:phijdot2}
}
which is larger than Eq.~(\ref{eq:phijdot}) by a factor of $\xi^{1/2}$. 
Thus, $\dot{\phi}_1^{}(t)$ indicates spike-like behavior around the origin. 
The corresponding time scale $\Delta t_{\text{sp}}^{0}:=1/m_{\text{sp}}^{0}$ is given by
\aln{\Delta t_{\text{sp}}^{0}=\frac{1}{m_{\text{sp}}^0}=\frac{M_{pl}^{}/(\sqrt{6}\xi)}{\dot{\phi}_1^{}}\sim \frac{1}{\sqrt{3\lambda_H^{}}M_{pl}^{}},
\label{eq:spike scale 1}
} 
which is quite small compared with the naive cutoff time scale $(M_{pl}^{}/\xi)^{-1}$.
\begin{figure}[t]
\begin{center}
\includegraphics[width=8cm]{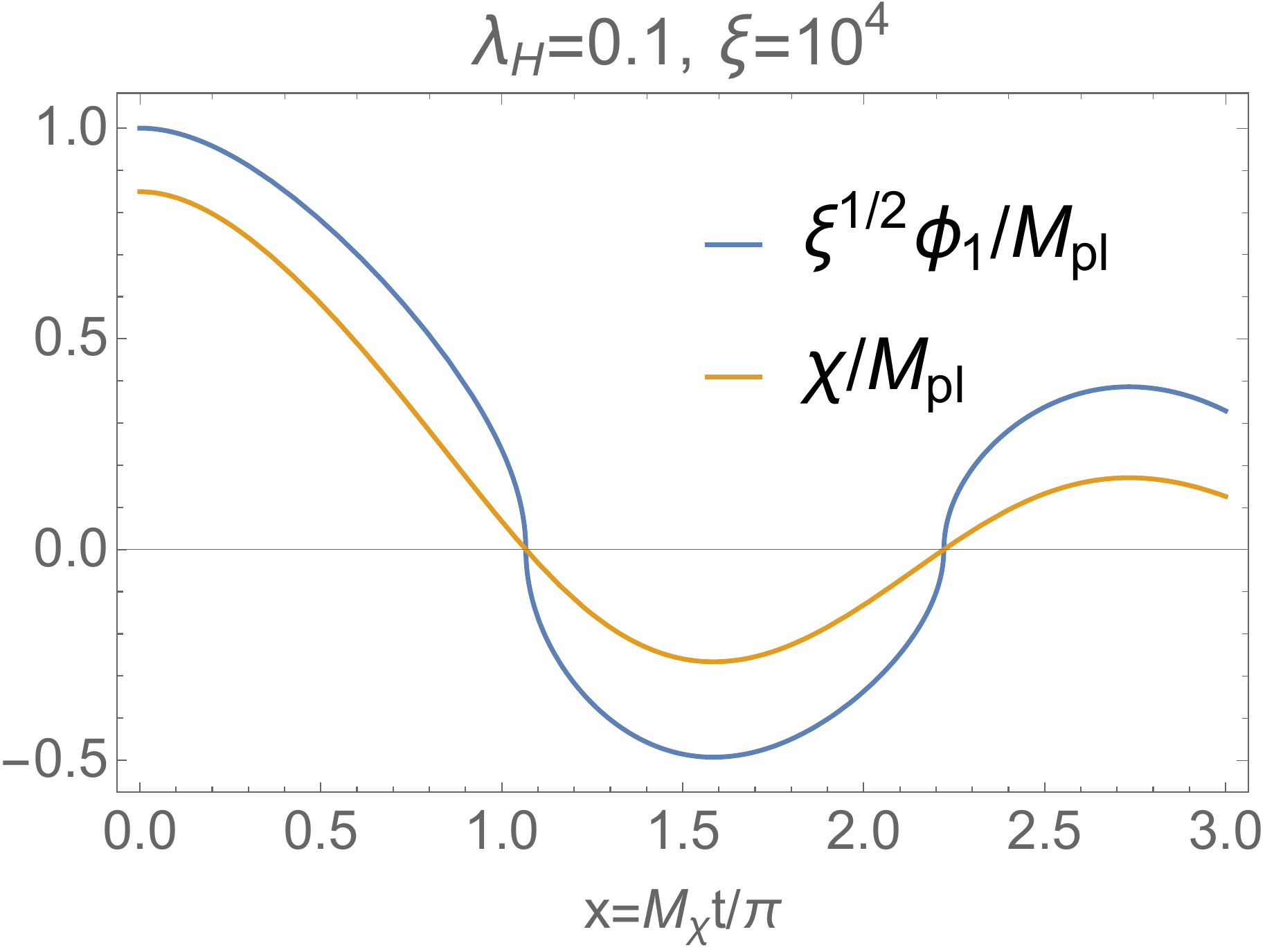}
\includegraphics[width=8cm]{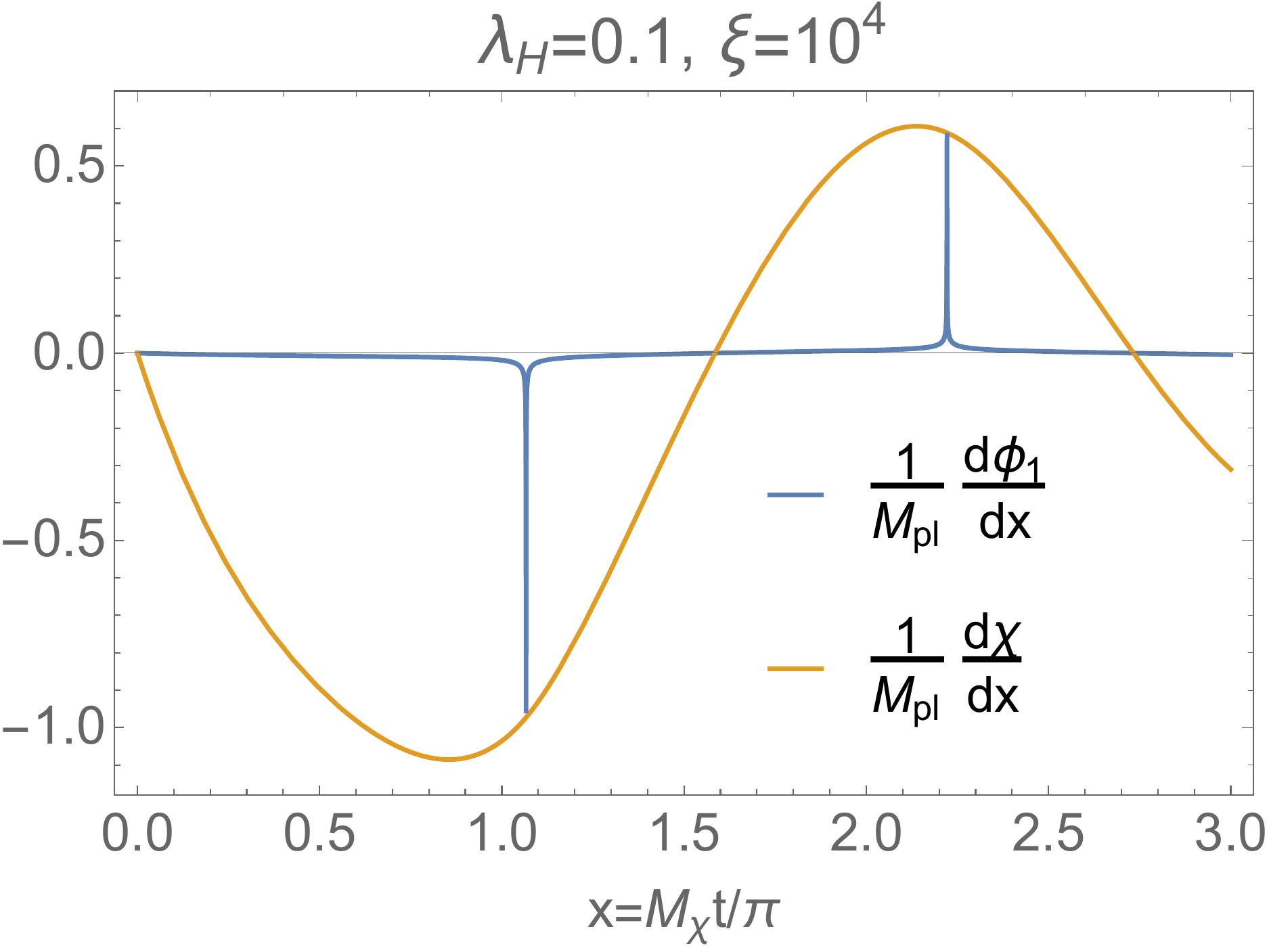}
\caption{Left: Time evolution of $\phi_1^{}$ (blue) and $\chi$ (orange). 
Right: Corresponding time evolution of $\dot{\phi}_1^{}$ (blue) and $\dot{\chi}$ (orange). 
We can see the spike-like behavior of $\dot{\phi}_1^{}$ around the first zero-crossing.  
}
\label{fig:evolution}
\end{center}
\end{figure}
In Fig.~\ref{fig:evolution}, we show our numerical calculations where blue (orange) lines correspond to $\phi_1^{}\ (\chi)$ and $x:= M_\chi^{}t/\pi$.   
In the left (right) panel, we show $\phi_1^{}\ (\dot{\phi}_1^{})$. 
One can in fact see that $\dot{\phi}_1^{}$ shows a spike-like behavior around the first zero crossing and its peak value is consistent with Eq.(\ref{eq:phijdot2}).  
This behavior of $\dot{\phi}_1^{}(t)$ was first pointed out in \cite{Ema:2016dny}, and this property plays an important role in particle production after inflation when the mass of the particles depends on the derivatives of $\phi_1^{}$. 

For completeness, we present the equation of motion of $\phi_1^{}$. 
This can be obtained by rewriting Eq.~(\ref{eq:eom fo varphi}) using Eq.~(\ref{eq:relation between varphiJ and varphi});
\aln{
&\ddot{\phi}_1^{}+3{\cal H}\dot{\phi}_1^{}+\frac{d^2\chi}{d\phi_1^2}\frac{d\phi_1^{}}{d\chi}\dot{\phi}_1^2+{\left(\frac{d\phi_1^{}}{d\chi}\right)^2}\frac{\partial V^{}}{\partial \phi_1^{}}=0.\label{eq: conventional eom}
}

From the above equation of motion and Eq.~\eqref{eq:phijdot2}, one can see 
\aln{
&{d^{2n}\phi_1 \over d t^{2n}}\sim \paren{\sqrt{\lambda_H} M_{pl}}^{2n} \phi_1, 
&&{d^{2n+1}\phi_1 \over d t^{2n+1}}\sim \paren{\sqrt{\lambda_H} M_{pl}}^{2n+1} {M_{pl}^{} \over \xi}, \quad \quad \text{for  $0\leq n\in \mathbb{Z}$},
}
for the first zero-crossing. Here we have neglected the Hubble friction term. 

\section{NG Mode}\label{app:NG}
Here, we present the derivation of the equation for the NG mode. 
Assuming the Friedmann metric, the kinetic term of $\theta_J^{}$ becomes 
\aln{
\frac{1}{2}\int d\eta\, d^3x\, F^2\bigg(\theta_J'^2-(\nabla\theta_J)^2\bigg)
=\frac{1}{2}\int d\eta\, d^3x\left({\theta'}^2-(\nabla\theta)^2+\frac{F''}{F}\theta^2\right),
\label{eq:kinetic term of theta}
}
where
\aln{
&F:={a r_J\over\Omega M_{pl}} \sqrt{1+ {c \,\Omega^2 \dot{r}_J^2 \over \Lambda_J^4}},
&&\theta:=F\theta_J.
}
The frequency $\omega_k$ is given by Eq.~\eqref{NG effective mass}.
The background $r_J$ is determined by
\aln{\label{Eq:r_J}
&\ddot{r}_J\sqbr{1+ {3c\over\Lambda_J^4} \dot{r}_J^2 \paren{dr_J\over dr}^2}+3{\cal H}\dot{r}_J\sqbr{1+ {c\over \Lambda_J^4} \,\dot{r}_J^2 \paren{dr_J\over dr}^2}+\frac{d^2r}{dr_J^2}\frac{dr_J}{dr}\dot{r}_J^2
-3\left(\frac{dr_J}{dr}\right)^2\frac{c\frac{{\partial\Lambda}_J}{\partial r_J}}{\Lambda_J^5}\dot{r}_J^4
+{\left(\frac{dr_J^{}}{dr}\right)^2}\frac{\partial V}{\partial r_J}=0,
}
where
\aln{
\frac{dr}{dr_J}=\frac{\sqrt{1+\xi(1+6\xi)\frac{r_J^2}{M_{pl}^2}}}{1+\frac{\xi r_J^2}{M_{pl}^2}}.
}

\bibliography{Bibliography}\bibliographystyle{utphys}
\end{document}